\documentclass[draft]{agujournal2019}
\usepackage{url} %this package should fix any errors with URLs in refs.
\usepackage{lineno}
\usepackage[finalnew]{trackchanges} %for better track changes. finalnew option will compile document with changes incorporated.
\usepackage{soul}
\usepackage{gensymb}
\draftfalse

\journalname{JGR: Space Physics}

\begin{document}

%%%%%%%%%%%%%%%%%%%%%%%%%%%%%%%%%%%%%%%%%%%%%%%
%  TITLE
%
% (A title should be specific, informative, and brief. Use
% abbreviations only if they are defined in the abstract. Titles that
% start with general keywords then specific terms are optimized in
% searches)
%
%%%%%%%%%%%%%%%%%%%%%%%%%%%%%%%%%%%%%%%%%%%%%%%

% Example: \title{This is a test title}

\title{Auroral Acceleration Generates Electron Beams in Jupiter's Middle Magnetosphere}

%%%%%%%%%%%%%%%%%%%%%%%%%%%%%%%%%%%%%%%%%%%%%%%
%
%  AUTHORS AND AFFILIATIONS
%
%%%%%%%%%%%%%%%%%%%%%%%%%%%%%%%%%%%%%%%%%%%%%%%

% Authors are individuals who have significantly contributed to the
% research and preparation of the article. Group authors are allowed, if
% each author in the group is separately identified in an appendix.)

% List authors by first name or initial followed by last name and
% separated by commas. Use \affil{} to number affiliations, and
% \thanks{} for author notes.
% Additional author notes should be indicated with \thanks{} (for
% example, for current addresses).

% Example: \authors{A. B. Author\affil{1}\thanks{Current address, Antartica}, B. C. Author\affil{2,3}, and D. E.
% Author\affil{3,4}\thanks{Also funded by Monsanto.}}

\authors{June Piasecki\affil{1}, Joachim Saur\affil{1}, George Clark\affil{2}, Barry H. Mauk \affil{2}, Annika Salveter\affil{1}, Jamey Szalay\affil{3}}

\affiliation{1}{Institute of Geophysics and Meteorology, University of Cologne, Cologne, Germany}
\affiliation{2}{The Johns Hopkins University Applied Physics Laboratory, Laurel, MD, USA}
\affiliation{3}{Department of Astrophysical Sciences, Princeton University, Princeton, NJ, USA}

% \affiliation{4}{Fourth Affiliation}

%\affiliation{=number=}{=Affiliation Address=}
%(repeat as many times as is necessary)

% Corresponding author mailing address and e-mail address:

% (include name and email addresses of the corresponding author.  More
% than one corresponding author is allowed in this LaTeX file and for
% publication; but only one corresponding author is allowed in our
% editorial system.)

% Example: \correspondingauthor{First and Last Name}{email@address.edu}

\correspondingauthor{June Piasecki}{jpiasec1@uni-koeln.de}

%%%%%%%%%%%%%%%%%%%%%%%%%%%%%%%%%%%%%%%%%%%%%%%
% KEY POINTS
%%%%%%%%%%%%%%%%%%%%%%%%%%%%%%%%%%%%%%%%%%%%%%%
%  List up to three key points (at least one is required)
%  Key Points summarize the main points and conclusions of the article
%  Each must be 140 characters or fewer with no special characters or punctuation and must be complete sentences

% Example:
% \begin{keypoints}
% \item	List up to three key points (at least one is required)
% \item	Key Points summarize the main points and conclusions of the article
% \item	Each must be 140 characters or fewer with no special characters or punctuation and must be complete sentences
% \end{keypoints}
 %electron beams within 30–1,200 keV are fitted with beam intensity function to qualitatively analyze their characteristics
%TC:ignore
\begin{keypoints}
\item Field-aligned electron beams are present throughout the equatorial middle magnetosphere within 14 and 50 $\mathrm{R_J}$ radial distance
\item Based on occurrence and energy fluxes, these beams are consistent with originating from auroral upward accelerated electrons
\item Most of the beam electrons are scattered out of the loss cone and serve as an energetic electron source in the middle magnetosphere
\
\end{keypoints}
%TC:endignore
%%%%%%%%%%%%%%%%%%%%%%%%%%%%%%%%%%%%%%%%%%%%%%%
%
%  ABSTRACT and PLAIN LANGUAGE SUMMARY
%
% A good Abstract will begin with a short description of the problem
% being addressed, briefly describe the new data or analyses, then
% briefly states the main conclusion(s) and how they are supported and
% uncertainties.

% The Plain Language Summary should be written for a broad audience,
% including journalists and the science-interested public, that will not have 
% a background in your field.
%
% A Plain Language Summary is required in GRL, JGR: Planets, JGR: Biogeosciences,
% JGR: Oceans, G-Cubed, Reviews of Geophysics, and JAMES.
% see http://sharingscience.agu.org/creating-plain-language-summary/)
%
%%%%%%%%%%%%%%%%%%%%%%%%%%%%%%%%%%%%%%%%%%%%%%%
%% \begin{abstract} starts the second page
\begin{abstract}
Observations made by the Juno spacecraft above Jupiter's polar regions have revealed that electrons accelerated toward Jupiter, which contribute to auroral emissions, are frequently accompanied by electrons accelerated away from Jupiter.
These electrons should be observable as narrow electron beams in the middle magnetosphere, in accordance with the principles of adiabatic particle motion. The existence of such beams has been previously reported using data from the Galileo mission, and their relation to auroral processes has been hypothesized.  
In the present study, we analyze electrons measured by Juno's JEDI instrument in the middle magnetosphere between 13 $ \mathrm{R_J}$ and 50.5 $\mathrm{R_J}$ radial distance and within energies of 30–1,200 keV.
The pitch angle distributions of potential electron beams are fitted with an intensity 'beamness' function. The presence of narrow beams is demonstrated throughout the observation range. 
The energy fluxes of auroral and equatorial electron beams are compared by including pitch angle scattering processes along the magnetospheric field lines. This is achieved by solving the pitch angle diffusion equation for different sets of diffusion coefficients.
The statistical occurrence distribution and the energy fluxes of the beams are consistent with auroral upward accelerated electrons observed in studies of the polar space environment. This finding provides further support for the hypothesis that the electron beams observed in the middle magnetosphere originate from the auroral acceleration region.
\end{abstract}
%TC:ignore
\section*{Plain Language Summary}

Jupiter has the most powerful aurora in the solar system, which is currently studied by NASA's Juno spacecraft.
One of the Juno mission's discoveries was that electrons above Jupiter’s auroral regions are accelerated not only toward the planet, where they can generate auroral emissions, but also in the opposite direction. As these electrons propagate along magnetic field lines, they should be observable in the middle magnetosphere, which is magnetically connected to Jupiter's auroral region.
These electrons are expected to appear as magnetic field-aligned distributions called electron beams.
Previously, such electron beams were observed in the magnetosphere using data from the Galileo mission. It was theorized that these electrons might originate from Jupiter's auroral regions.
We analyze electrons measured by Juno in the middle magnetosphere to detect electron beams and further confirm this hypothesis.
We calculate different properties of the electron beams and compare them with auroral electrons observed above Jupiter's polar region to determine the relation between both electron populations.
We find that the occurrence distribution and the energy flux of the electron beams are similar to those of the auroral electrons.
These results further support the hypothesis that electron beams in the middle magnetosphere originate from the auroral region.
%TC:endignore

%Enter your Plain Language Summary here or delete this section.
%Here are instructions on writing a Plain Language Summary: 
%https://www.agu.org/Share-and-Advocate/Share/Community/Plain-language-summary
%%%%%%%%%%%%%%%%%%%%%%%%%%%%%%%%%%%%%%%%%%%%%%%
%
%  BODY TEXT
%
%%%%%%%%%%%%%%%%%%%%%%%%%%%%%%%%%%%%%%%%%%%%%%%
%%% Suggested section heads:
% \section{Introduction}
%
% The main text should start with an introduction. Except for short
% manuscripts (such as comments and replies), the text should be divided
% into sections, each with its own heading.
% Headings should be sentence fragments and do not begin with a
% lowercase letter or number. Examples of good headings are:
% \section{Materials and Methods}
% Here is text on Materials and Methods.
%
% \subsection{A descriptive heading about methods}
% More about Methods.
%
% \section{Data} (Or section title might be a descriptive heading about data)
%
% \section{Results} (Or section title might be a descriptive heading about the
% results)
%
% \section{Conclusions}
%was muss rein:  discussuion radial diffusion abgrenzung wi ebei nenon, clark 2014 zitieren
%Text here ===>>>
% previous work:\\
% equatorial electron beams (thomas, frank and paterson, mauk and saur)\\
% juno observations aurora  ( mauk 2017,2018,2020, clark2017, allegrini2017, salveter 2022, sulaiman2022, connerney 2018,2022\\
% scattering elliot 2018, ..)\\
% hypothesis and structure of the work\\
\section{Introduction}
Jupiter's strong magnetic field, its high rotation rate, and the internal plasma source Io create a unique magnetospheric system that generates the powerful aurora of Jupiter, which is currently being studied by the Jupiter orbiter Juno \cite{bolton2017}.
One of the great discoveries of the Juno mission were bidirectional electron beams over Jupiter’s polar regions commonly associated with the main auroral emission zone \cite{mauk2017a}. Previous observations by the Galileo spacecraft found electron beams in the middle magnetosphere of Jupiter, which were hypothesized to originate from the auroral acceleration regions close to Jupiter \cite{frank2002}. In this study we analyze the middle magnetosphere electron beams using Juno observations to confirm whether these beams originate from the upward accelerated auroral electrons.

Galileo observations revealed the presence of bidirectional electron beams in the equatorial middle magnetosphere of Jupiter, describing magnetic field-aligned electron distributions with high intensities parallel and anti-parallel to the magnetic field direction, within energies of $1 -50~\mathrm{keV}$ \cite{frank2002} and $29 -884~\mathrm{keV}$ \cite{tomas2004a,tomas2004b}.
The electron measurements revealed a change in the intensity-pitch angle distributions from signatures of trapped electron distributions, i.e. intensity maximum at $90\degree$ pitch angle, to dominating bidirectional beam distributions, i.e. intensity minimum at $90\degree$ pitch angle, at distances between 10 and 17 $\mathrm{R_{J}}$ \cite{tomas2004a}.
Similar observations have been reported for electrons in the magnetosphere of Saturn \cite{saur2006,clark2014}. 
\citeA{mauk2007} introduced a qualitative analysis of electron beams to investigate their characteristics and reported the occurrence of narrow beams interrupted by wider scattered beams and isotropic distributions for electron energies within $15 -42~\mathrm{keV}$.
%(for 29 to 884 keV electrons.) 
It has been suggested that the middle magnetosphere beams may be associated with auroral processes, since this magnetosphere region maps magnetically to the aurora \cite{frank2002,tomas2004a,tomas2004b,mauk2007}. 
%However, the mechanism that creates these energetic field-aligned distributions remains unclear. 
Electron acceleration at low altitudes in the auroral acceleration region is proposed as a possible heating mechanism \cite{frank2002,mauk2007}.
An isotropic electron distribution that was energized at low altitudes would exhibit a typical narrow beam width in the equatorial middle magnetosphere, considering the conservation of the first adiabatic invariant \cite{frank2002}. 
\citeA{mauk2007} discuss that upward acceleration of electrons in the main auroral regions contradicts the traditional expectations of electrostatic planetward acceleration in the region that is thought to be dominated by upward field-aligned currents of the magnetospheric large-scale current system \cite{cowleybunce2001,hill2001,knight1973}.
Other processes have been discussed as potential mechanisms for generating field-aligned electron distributions, including the scattering of electrons to smaller pitch angles \cite{tomas2004b} and the outward adiabatic transport of energized electrons \cite{woodfield2014}.

\begin{figure}
\centering
\noindent\includegraphics[width=0.9\textwidth]{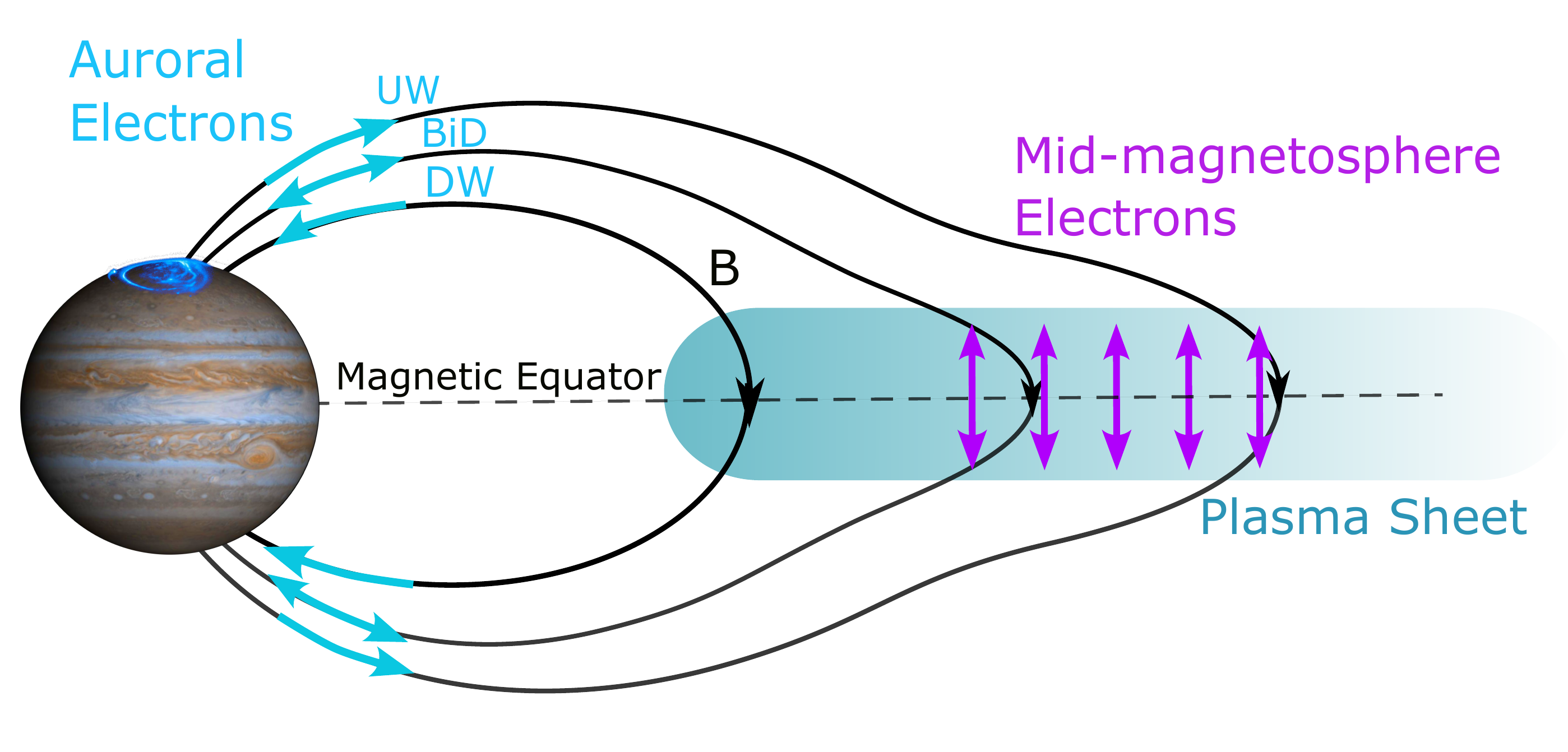}
\caption{A schematic of the Jovian magnetosphere. The field-aligned electrons measured above the main aurora are represented by blue arrows based on Figure 9 from \citeA{mauk2020}. They describe upward, bidirectional and downward accelerated electrons, organized according to the results from \citeA{salveter2022}. The bidirectional electron beams measured in the middle magnetosphere are shown as purple arrows.}
\label{magnetosphere}
\end{figure}

% % juno electron observations
Later observations of electrons ($0.1 -1,000~\mathrm{keV}$) made by the Jupiter polar orbiter Juno confirmed the transition from trapped electron distributions for $M\leq 15$ to bidirectional and isotropic distributions, which are present up to distances of $M=80$ \cite{ma2021}. The $M$-shell parameter $M$ is defined as the jovicentric radial distance at which a modeled magnetic field line intersects the magnetic equator, normalized by the planet radius, as described by \citeA<e.g.>{ma2021}.
Moreover, the arrival of Juno enabled detailed studies of auroral acceleration processes through in situ measurements directly above the Jovian aurora.
\citeA{mauk2017a} reported for energetic electrons ($30-1,000~ \mathrm{keV}$) that upward accelerated electrons are observed in the auroral acceleration region, in addition to the electrons accelerated toward Jupiter, which were expected for the auroral processes. These upward-going electrons, which are known to propagate towards the magnetosphere, were found to show similarities with the equatorial electrons previously analyzed by \citeA{mauk2007}. 
The corresponding energy spectra are frequently broad, indicating that these electrons are associated with a stochastic acceleration process \cite{mauk2017a,mauk2017b,saur2018,clark2018}. 
Further measurements revealed the predicted existence of upward field-aligned electric potentials and electron energy spectra with mono-energetic features suggesting electrostatic acceleration \cite{mauk2017a, clark2017,clark2018}. However, in contrast to previous expectations, the highest downward energy fluxes were observed to originate from broadband electron distributions \cite{mauk2017b}.
% % zones
Based on the repeatable characteristics of the auroral electrons, \citeA{mauk2020} divided the main aurora into three distinct zones, organized from lower to higher latitudes:
(1) The diffuse aurora, without signatures of active electron acceleration but characterized by electrons scattered to field-aligned pitch angles, where they can reach the atmosphere. (2) Zone I is associated with downward accelerated electrons and (3) Zone II is associated with upward accelerated or bidirectional electron distributions.
%This suggests that an acceleration mechanism is involved that can generate bidirectional electron distributions.%, contradicting the previous suggested acceleration through electrostatic potentials.
\citeA{salveter2022} examined the statistical occurrence and energy fluxes of energetic electrons that represent the defined zones above the main aurora. 
They showed that downward-going electrons show a high occurrence above $50\%$ of all field-aligned distributions at smaller magnetic latitudes below $\theta=78\degree$, while upward-going electrons dominate for larger magnetic latitudes. In addition, bidirectional electrons show the largest occurrence in between.
However, all distribution types are observed throughout the observation range. This demonstrates the persistent occurrence of upward accelerated electrons in the auroral acceleration regions, which may serve as a source of middle magnetosphere beams.
The positions and characteristics of the auroral electrons and the middle magnetosphere electron beams are displayed in the schematic of the Jovian magnetosphere in Figure \ref{magnetosphere}.

In this study we investigate the hypothesis that the auroral upward accelerated electrons, measured close to Jupiter in the auroral acceleration regions, are observable as narrow beams in the middle magnetosphere, as the electrons propagate along the magnetic field toward the other hemisphere.
We analyze JEDI electron pitch angle distributions within radial distances of $13~\mathrm{R_J}$ and $50.5~\mathrm{R_J}$ within energies of $30-1,200~ \mathrm{keV}$. 
The enhanced temporal and angular resolution, together with the substantial number of JEDI measurements, now enables a statistical analysis of the middle magnetosphere beams. 
The objective of this study is to determine if the electron beams observed in the middle magnetosphere are, in fact, the auroral upward electrons, which would confirm the previous assumption that the beams are accelerated at auroral altitudes.
This is achieved through the quantitative characterization of electron beams and their energy fluxes.
The results are compared with the statistical analysis of auroral electrons by \citeA{salveter2022}.
It is assumed that pitch angle scattering is involved \cite{frank2002,mauk2007}, which broadens the angular width of the beams through wave-particle interaction, since their width would be less than a degree without any scattering. 
In Section \ref{data and methods} the data used in this study are described together with the methods developed for the investigation of electron beams.
The calculated occurrence and energy flux distributions of the beams are presented and analyzed in Section \ref{results} together with other properties of the beams. The study concludes with a summary of the main findings in Section \ref{conclusion}.
\section{Data and Methods}
\label{data and methods}
In this section, an overview of the JEDI data and the applied methods is provided. The electron classifications are organized according to three different spatial parameters, which are introduced in Section \ref{spatial parameters}. The qualitative criteria for the detection of beams are described in Section \ref{beam detection}. This is followed by a description of the beam characterization technique, which is adapted from the work of \citeA{mauk2007}. 
To estimate the energy fluxes of the beams, the evolution of the pitch angle distribution from their source region into the equatorial magnetosphere needs to be known. Therefore, we solve the pitch angle diffusion equation, as further explained in Section \ref{calculation of the energy flux}.

\subsection{Data} 
\begin{figure}[ht]
\centering
\noindent\includegraphics[width=0.8\textwidth]{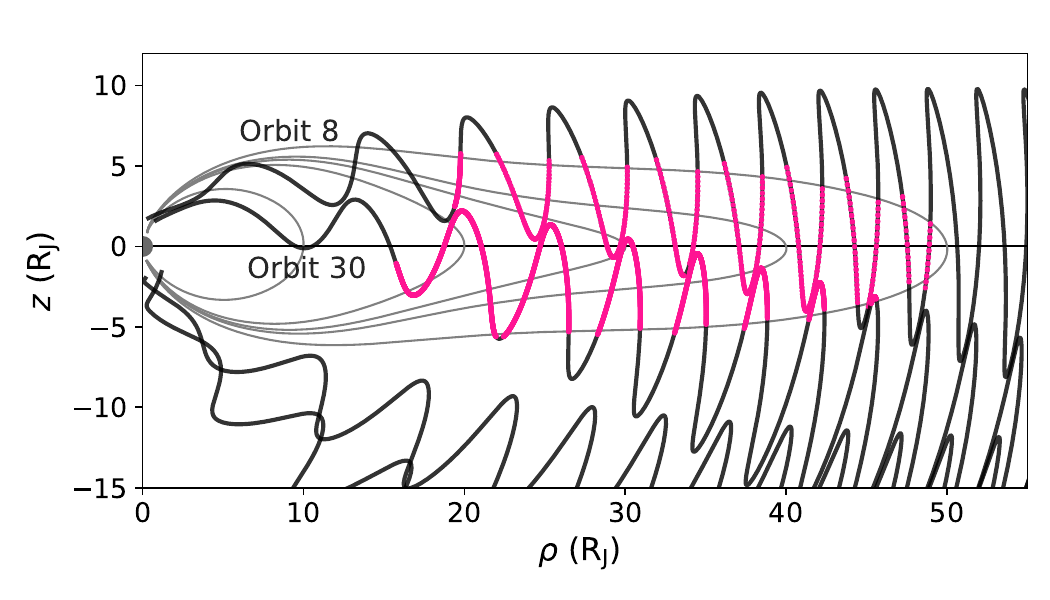}

\caption{Trajectory of Juno orbits 8 and 30 in the $\rho$-$z$ plane. The coordinate system of the trajectory is described by a tilt of the System III axes by the VIP4 \cite{connerney1998} dipole tilt of $9.5\degree$ towards longitude $201\degree$. The magnetic field lines are calculated with the JRM33 \cite{connerney2022} and CON2020 model \shortcite{connerney2020}. The data used in this analysis, which magnetically map to a maximum $M$-shell of $M=50.5$ are marked as pink orbit segments. 
}
\label{fig:orbits}
\end{figure}

The electron measurements that are the focus of this study were obtained by Juno's Jupiter Energetic particle Detector Instrument JEDI on the inbound trajectory of orbits 8 to 30 within radial distances of 13 $\mathrm{R_J}$ and 50.5 $\mathrm{R_J}$. Due to the magnetic dipole tilt, Juno traversed the equatorial magnetosphere on multiple occasions during each orbit.
Orbits 8 and 30 are displayed in Figure \ref{fig:orbits} in the $\rho$-$z$ plane of a magnetic dipole-oriented coordinate system, where $\rho$ is the distance from the magnetic axis and $z$ is the distance from the magnetic equator. For this plot, the magnetic coordinates of the trajectory are based on the VIP4 magnetic field model \cite{connerney1998}. The magnetic field lines are modeled using the Jovian internal magnetic field model JRM33 \cite{connerney2022} and the magnetodisc current sheet model CON2020 \shortcite{connerney2020} for the external magnetic field.
These models are calculated with the C++ Python wrapper of the community coding project \cite{wilson2023, James_JupiterMag}. 
The pink segments represent the measurement intervals that were selected for our study, which is described in more detail in Section \ref{spatial parameters}.

JEDI measures the energy and angular distributions of energetic electrons with energies ranging from $<40$ to $>500$ keV.
The instrument is composed of three sensors: The JEDI-90 and JEDI-270 sensors with six look directions over a $160\degree$ field of view (FOV), and JEDI-180 over a $148\degree$ FOV.
In each look direction, a solid-state detector measures the total energy of incoming particles.
The JEDI-90 and JEDI-270 sensors are installed approximately in the equatorial plane of the spacecraft, with the JEDI-180 instrument positioned perpendicular to them.
Whether a full angular view is obtained for a given measurement interval depends on the spacecraft's orientation relative to the magnetic field. 
The three sensors are arranged so that, close to Jupiter, the equatorial sensors provide nearly full pitch angle coverage almost instantaneously, whereas at further distances in the magnetosphere, the third sensor provides full coverage every 30 seconds (spin period)  \cite{jedi}. As this region is the subject of investigation in this study, data from the JEDI-180 instrument are used.
The calibrated data sets are available through the digital archive Planetary Data System \cite{jedidata} and comprise the differential intensity, also called intensity in units of counts per $\mathrm{cm}^2\cdot \mathrm{s}\cdot \mathrm{sr}\cdot\mathrm{keV}$.
For the region of our investigation, low-energy-resolution electron data are available, for which the measured energy range is divided into eight energy channels.
The efficiency with which the JEDI instruments measure electrons is found to significantly decrease and is not well characterized for energies below 30 keV \cite{mauk2017a}. Therefore, the two lowest energy channels are excluded from the calculations in this study.  \add{The low-energy-resolution electron data is available for radial distances $r\geq 13-20~\mathrm{R_J}$ for the orbits used in our study, which is why less data is available at smaller distances.}
The magnetic field measurement is performed by the Juno magnetic field investigation \cite{mag}, which is used to determine the pitch angle between the magnetic field direction and the velocity vector of the particle.

\subsection{Spatial Parameters} \label{spatial parameters}
The aim of our study is to verify a relation between the equatorial middle magnetosphere beams and the upward electrons measured at low polar auroral altitudes. In the polar study by \citeA{salveter2022} the auroral electrons are spatially organized according to the magnetic latitude and the dipole $L$-shell. Therefore, we also characterize the middle magnetosphere electrons according to spatial parameters, which are introduced in this section and allow for a better comparability. The dipole $L$-shell is defined as the planetocentric radial distance of a magnetic dipole field line in the magnetic equatorial plane, normalized by the planet radius \cite[p. 33]{baumjohann}.
Characteristics of auroral electrons organized from smaller to larger magnetic latitudes or $L$-shell should be represented in measurements of electron beams in the equatorial middle magnetosphere from smaller to larger distances, based on the geometry of the magnetic field and the first adiabatic invariant.
The challenge is to identify a suitable spatial parameter that is comparable and physically accurate to represent the electron beam distribution along the equatorial magnetosphere. We use three spatial parameters in our study, which are described in this section: the radial distance $r$, the $M$-shell $M$, and the \textit{polar} $L$-shell $L$.

% radal distanc
The radial distance $r$ is the only parameter that is independent of magnetic field models and is the least affected by uncertainties. Therefore, it can be used to provide a true spatial distribution that could approximate the polar auroral distribution as a first impression.
However, it is important to note that when electron measurements are organized by radial distance, the electron distributions that are assigned to the same radial distance may be measured at significantly different magnetic latitudes. Consequently, these measurements may magnetically map to different locations in the auroral acceleration region that are distant from each other.
%m shell

The $M$-shell parameter $M$ is defined as the radial distance at which a specific magnetic field line crosses the magnetic equator, the maximum radial distance of the magnetic field line from the planet, normalized by the planet radius. The definition is analogous to that of the dipole field $L$-shell, but considers a more realistic, non-dipolar magnetic field geometry that includes the influence of the plasma sheet and higher multi-pole moments. This parameter enables the mapping of electron beam positions along the field line to the magnetic equator. This provides a spatial characterization that is more comparable to the polar organization by magnetic latitude or $L$-shell.
The calculation of the $M$-shell parameter is based on the Jovian internal magnetic field model JRM33 \cite{connerney2022} combined with the magnetodisc current sheet model CON2020 \shortcite{connerney2020} for the external magnetic field.
%The C++ Python wrapper of the community coding project \cite{wilson2023, James_JupiterMag} is utilized to calculate these models. 
Due to the outer limit of the current sheet model of 51.4 $\mathrm{R_J}$, we restrict the measurements to observation positions that magnetically map to a maximum of $M = 50.5$. The resulting selection of measurement intervals can be seen exemplarily for Orbits 8 and 30 in Figure \ref{fig:orbits}. It should be noted that the current sheet model is based on magnetic field observations for $r\leq 30 ~\mathrm{R_J} $. Consequently, some larger deviations from the true magnetic field may be possible at larger distances. 
%l shell

Since in the comparison study of \citeA{salveter2022} the electron measurements are organized according to the $L$-shell, we introduce a spatial parameter called \textit{polar} $L$-shell.
This parameter is introduced to magnetically map the measurements in the middle magnetosphere of our study to positions within the polar observation region of \citeA{salveter2022}.
For this calculation we trace the middle magnetosphere measurement positions along the magnetic field line to the mean radial distance of polar Juno positions during the observations used by \citeA{salveter2022} of about 1.85 $\mathrm{R_J}$.
For this position, the $ L$-shell is then calculated similar to the calculation in \citeA{salveter2022}, using only the \add{dipolar moment of the }internal magnetic field model JRM33.
The idea is that a perfect tracing would assign electrons measured at the same magnetic field line, in the middle magnetosphere and the polar observation region, to the same $L$-shell and $polar~L$-shell, respectively.
However, the calculation is based on an axisymmetric external magnetic field model without local time dependence \cite{connerney2020}, for which deviations are expected for $r> 30 ~\mathrm{R_J} $. 
Furthermore, field-line tracing along such long distances from the magnetosphere to the polar space region of Jupiter is sensitive to small changes. Therefore, larger uncertainties are expected for this spatial parameter than for the calculated $M$-shell.
However, the objective is to improve the comparability of the two studies by providing an approximate positioning of our measurements within the observation range of \citeA{salveter2022}, rather than expecting a highly accurate calculation.
\subsection{ Beam Detection}\label{beam detection}
\begin{figure}[ht]
\centering
\noindent\includegraphics[width=0.8\textwidth]{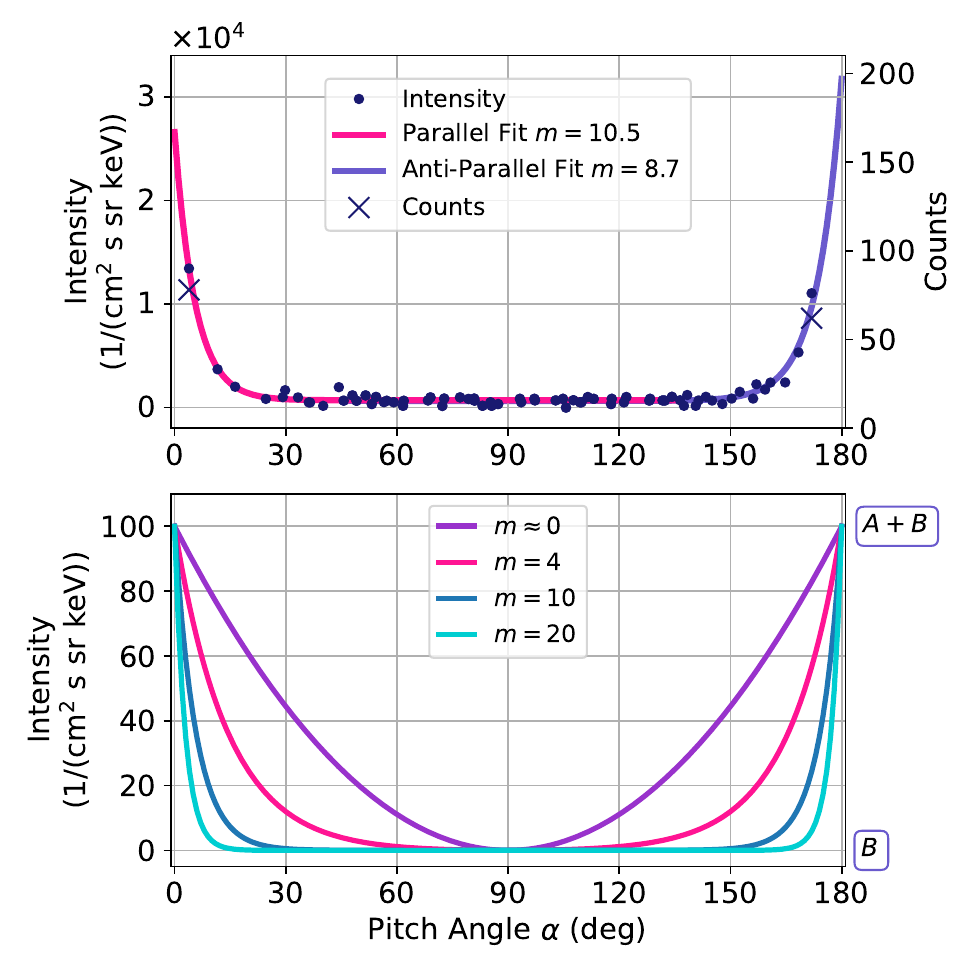}
\caption{(Top) An electron intensity-pitch angle distribution (dots) fitted with the beam intensity function (solid lines). The crosses are the counts for the two data points closest to the magnetic field direction. This distribution is measured in energy channel $E_2$ with central energy $\bar{E_2} =32 ~\mathrm{keV}$ on day 36 of 2018. (Bottom) The beam intensity function for different example beamness parameters $m$, with $A=100$ and $B=0$.}
\label{padexample}
\end{figure}

We developed a method to determine whether an electron pitch angle distribution includes a beam to provide an automated procedure that ensures reproducibility.
The upper plot of Figure \ref{padexample} presents an example intensity-pitch angle distribution $I(\alpha)$ of a beam, with the measured intensities represented as dots.
For a distribution to be classified as a beam, it must fulfill the following three criteria, which are tested for the measured intensity of each energy channel $E_i$ and each direction, parallel and anti-parallel to the magnetic field, separately. 

(1) Sufficient pitch angle coverage: The minimum measured angle with respect to the magnetic field direction must be less than $14\degree $, which is approximately half the FOV of the JEDI sensors and should ensure the possibility to measure a beam. This results in minimum measured pitch angles of $\alpha < 14\degree$ and $\alpha>166\degree$ for the parallel and anti-parallel directions, respectively.

(2) Enhanced field-aligned intensity: This criterion ensures that the distribution exhibits higher intensities in the field-aligned direction than around $\alpha=90 \degree$, which characterizes the typical high beam intensity compared to the isotropic background electron intensity.
We decided that the intensity data points closest to the magnetic field direction, $I(\alpha_{min})$ for the parallel and $I(\alpha_{min'})$ for the anti-parallel direction, respectively, must exceed six standard deviations $\sigma_{\mathrm{iso}}$ above the mean intensity of the isotropic background electrons $I_{\mathrm{iso}}$ to be considered a possible beam:
\begin{linenomath*}
\begin{equation}
I(\alpha_{min}) \geq 6\times \sigma_{\mathrm{iso}} + I_{\mathrm{iso}}
\end{equation}
\end{linenomath*}
The standard deviation $\sigma_{\mathrm{iso}}$ and the mean intensity $I_{\mathrm{iso}}$ are calculated for the electron intensities $I_n$ with $n = 1,...,N$ measured within pitch angles of $45\degree -135\degree $ according to
\begin{linenomath*}
\begin{equation}
I_{\mathrm{iso}}=\frac{1}{N}\sum_n I_n ~~~~~~~~~~\sigma = \sqrt{\frac{1}{N}\sum_n(I_n-I_{\mathrm{iso}})^2}
\end{equation}
\end{linenomath*}
(3) Sufficient signal-to-noise ratio: It can be assumed that the electron measurements are Poisson-distributed events with statistical uncertainties, which are described by a signal-to-noise ratio of $SNR=\sqrt{n}$, where $n$ is the number of counts in the considered time interval \shortcite{salveter2022,bevington1993}.
For a beaming signal to be considered valid, a minimum of nine electron counts must be measured for the data point closest to the magnetic field, ensuring a signal-to-noise ratio of at least three.

%\subsection{Beam Characterization Technique}  \label{Beam Characterization Technique}
For further characterization of the selected beams, the technique of \citeA{mauk2007} is used, which involves the fitting of the pitch angle distributions with a beam intensity function of the following form
\begin{linenomath*}
\begin{equation}
    I_{Beam}(\alpha) = B+A~{ \cdot~ \frac{\cosh (m\cdot(\alpha-90)\cdot \frac{\pi}{180})-1}{\cosh(m\cdot\frac{\pi}{2})-1}}
    \label{eq:beam}
\end{equation}
\end{linenomath*}
where $\alpha$ is the pitch angle in degrees and $A, B$ and $m$ are the free fitting parameters.
As the value of $m$ increases, the beam becomes narrower, which is demonstrated for different values of $m$ in the bottom plot of Figure \ref{padexample}. Conversely, as $m$ approaches zero, the function approaches a parabola.
This \textit{beamness }parameter $m$ allows for the differentiation between true narrow beams and 'scattered beams', which are characterized by a greater width. \citeA{mauk2007} determined that distributions with a beamness of $m \geq 4$ are classified as true narrow beams, while those with $0 \leq m < 4 $ are considered scattered beams. 
The narrow beam distributions, characterized by a sharp intensity reduction from the magnetic field direction towards $90\degree$, clearly differentiate the beam intensity from the intensity of the more isotropic background electrons.
The parameter $B$ corresponds to the minimum intensity at $\alpha = $$90\degree$ and characterizes the intensity of the isotropic background electrons. The parameter $A$ defines the intensity of the beam above the isotropic background intensity exactly in the direction of the magnetic field, resulting in a total differential intensity of $A+B$ at pitch angles of $0\degree$ and $180\degree$, as demonstrated in Figure \ref{padexample}. 
In addition to the selection of narrow beams, this method enables the estimation of the beam's intensity at pitch angles close to the magnetic field direction, which are often not well covered by JEDI.

For the purposes of this study, we adapted the beam fitting method of \citeA{mauk2007} as follows.
To resolve asymmetries, the fit is performed separately for parallel and anti-parallel beams in the angular range $0^{\circ} \leq \alpha \leq 135^{\circ}$ and $45^{\circ} \leq \alpha \leq 180^{\circ}$, respectively.  The pitch angle limits are set beyond the $90\degree$ midpoint to ensure the best possible representation of the intermediate isotropic distribution for both fitting directions.
Figure \ref{padexample} shows the fits of the example beam distribution as solid lines. 

The JEDI intensity data points do not show a measurement at a specific pitch angle position, but rather an integration of measured electrons over a pitch angle range. The intensity is determined for observations within the FOV of the JEDI sensors, which have an instantaneous full-width-at-half-maximum FOV of about $9\degree \times 17 \degree$.
Therefore, we do not evaluate the analytic beam intensity function, Equation  (\ref{eq:beam}), at each data pitch angle position in the fitting algorithm, but instead apply a weighting scheme across a pitch angle range to more accurately reflect the actual measurement process.
 Before fitting, the beam intensity function is weighted using a Gaussian function with a height of 1 and an FWHM of 9° in the range of $\pm 4.5\degree$ in pitch angle for each data position.

The quality of the fit depends on the pitch angle coverage of each fitted data interval and is occasionally supported by only a few data points close to field-aligned directions. 
An optimized fit can be validated by analyzing the minimized cost function for varying values of the fitting parameters.
A successful fit yields optimized parameters that correspond to the global minimum of the cost function within the parameter space.
The Appendix provides figures of the cost function space analysis of two example measurement intervals, calculated similarly to the method described in \citeA{kim2020}.
These results show that the best-fit parameters result in a minimum cost function value, demonstrating a successful minimization.
The analysis reveals a positive correlation between the fitting parameters $A$ and $m$, with the degree of correlation being influenced by the pitch angle coverage. 
As the parameters depend on each other and the precision with which the best-fit parameters can be determined depends on pitch angle coverage, which varies for all measurements, we intend to consider fitting parameter errors in our analysis.
In the differentiation between narrow beams ($m\ge4$) and scattered beams ($m<4$) based on their beamness parameter $m$, we incorporate the parameter error $\sigma_m$, ensuring that beams that cannot be clearly assigned to a category, because of a large parameter uncertainty, are excluded from that category. In our results, a beam is then defined as a true narrow beam if $m - \sigma_m \ge 4$ and as a scattered beam if $m + \sigma_m < 4$. 
Parameter errors are further considered in the calculation of the average energy fluxes, as described in the following Section \ref{calculation of the energy flux}.

\subsection{Calculation of the Energy Flux}\label{calculation of the energy flux}

For time intervals for which narrow beams are detected in all energy channels, the energy flux is calculated in order to compare it with the energy flux of the auroral electrons studied by \citeA{salveter2022}.
The energy flux equation as a function of intensity or differential particle flux $I(\alpha, E)$ can be derived from the particle flux through a surface perpendicular to the magnetic field direction \cite[p. 121]{baumjohann}. The particle flux can be described by 
\begin{linenomath*}
\begin{equation}
F_N=\int_{\theta_{min}}^{\theta_{max}} d\theta\int_{\alpha_{min}}^{\alpha_{max}} d\alpha\int_{E_{min}}^{E_{max}} dE~I(E,\alpha ) \cos\alpha~\sin\alpha
\end{equation}
\end{linenomath*}\cite[p. 121, 125]{baumjohann}.\remove{where $f(\mathbf{v})$ is the phase space distribution function, $v_\perp=v\sin\alpha$ is the perpendicular, and $v_\parallel=v\cos{\alpha} $ is the parallel velocity component of the particles.
The intensity $I(E,\alpha)$ as a function of energy can be related to the distribution function by $f(\mathbf{v},\alpha )~v^3~dv~d\Omega = I(E,\alpha)~dE~d\Omega$. In this study, the fit of $I_{Beam}( \alpha )$ corresponding to Equation  (3) is used for the calculation of the energy flux within a specific energy channel $E_i$ with central energy $\bar{E_i}$ and the width of the energy channel $\Delta E_i$ . }
\add{The differential energy flux can then be obtained by multiplying the intensity by the particle energy. 
Since the intensity considered in our study is measured and fitted within specific energy channels $E_i$ within the limits $E_{i,min}$ and $E_{i,max}$, these particles are assigned to the central energy of that channel $\bar{E_i}$. 
The calculation of the energy flux of each energy channel is simplified to}
\begin{linenomath*}
\begin{equation}
    F_{E_i} = \bar{E_i}\cdot \Delta E_i \int_{\theta_{min}}^{\theta_{max}} d\theta\int_{\alpha_{min}}^{\alpha_{max}} d\alpha~I_{Beam,i}( \alpha )  ~\cos\alpha~\sin\alpha
    \label{eq:singlefe}
\end{equation}
\end{linenomath*}
\add{ where $I_{Beam,i}( \alpha )$ is the beamness function fit of the energy channel $E_i$ corresponding to Equation} (\ref{eq:beam}). The central energy $\bar{E_i}$ and the width of the energy channels $\Delta E_i$ are calculated by
 \begin{linenomath*}
\begin{eqnarray}
     &&\bar{E_i}=\sqrt{E_{i,min}\cdot E_{i,max}}\\
     &&\Delta E_i = E_{i,max}- E_{i,min}
 \end{eqnarray}
 \end{linenomath*}
The isotropic background intensity, defined by the fitting parameter $B_i$, is subtracted from the intensity fit to obtain the intensity of the beam without the background electrons.
The resulting intensity is integrated over the entire azimuthal angle assuming gyrotropicity. Accordingly, the energy flux is given by
 \begin{linenomath*}
 \begin{equation}
\label{eq:energyflux}
      F_{E_i} =2\pi\cdot \bar{E_i}\cdot \Delta E_i \int_{\alpha_{min}}^{\alpha_{max}} d\alpha~A_i~{ \cdot~ \frac{\cosh (m_i\cdot(\alpha -\frac{\pi}{2}) )-1}{\cosh(m_i\cdot\frac{\pi}{2})-1}}  ~\cos\alpha~\sin\alpha
 \end{equation}
\end{linenomath*}
where $m_i$, and $A_i$ are the fitting parameters of the electron pitch angle distribution of channel $E_i$. The total energy flux $F_E$ can be calculated as the sum of the energy fluxes $F_{E_i}$ of each channel $F_E = \sum  F_{E_i}$. 
Finally, Equation \ref{eq:energyflux} is integrated from $\alpha_{min}=0\degree$ to an integration limit $\alpha_{max}$, which is determined using different approaches.
The first approach is to include the whole beam intensity, which serves as an upper limit for the energy flux. 
The integration boundary $\alpha_{max}$ is simply set to $90\degree$ as the intensity contribution at pitch angles around $90\degree$ is negligible for the narrow beams considered with $m \geq 4$. This can be seen in Figure \ref{padexample}, which shows that the intensity drops rapidly towards $\alpha=90\degree$.
The second approach is to include only the intensity within the local loss cone at the measurement position in the middle magnetosphere. The intensity of the upward auroral electrons is measured within the polar local loss cone in \citeA{salveter2022}. Therefore, this integration boundary would correspond approximately to the width of the auroral beam in the absence of pitch angle scattering, since the electrons would remain inside the loss cone while propagating toward the magnetosphere.
The third approach involves the estimation of the scattering process to determine the width of a scattered beam as the integration boundary. This approach is further described in the following Section \ref{pitch angle diffusion}.
%Pitch angle scattering of electron beams was observed by \citeA{elliott2018a} for beams in the Jovian polar cap, the region poleward of the main aurora.

Several studies, including \citeA{salveter2022}, estimate the energy fluxes of measured electrons at Jupiter's upper atmosphere, where the auroral emissions are generated.
In our study, the total energy flux $F_{E_{Juno}}$ at the Juno measurement position is projected onto Jupiter's upper atmosphere by multiplying the value by the ratio of the magnetic field strength at the atmosphere and at the Juno measurement position in the middle magnetosphere
\begin{linenomath*}
\begin{equation}
F_{E_{atm}} =F_{E_{Juno}}\cdot\frac{B_{atm}}{B_{Juno}}
\end{equation}
\label{eq:projection}
\end{linenomath*}
This estimation assumes that electrons stay within a flux tube and uses the conservation of magnetic flux $\Phi=\mathbf{A\cdot B}$, where $\mathbf{A}$ is the cross-sectional area of a flux tube carrying the electrons and $\mathbf{B}$ is the magnetic field strength. 
Given that the measured intensity and the resulting energy flux are defined per unit area, it follows that the same electron distribution propagating through a flux tube will exhibit a higher intensity close to the planet in a narrow flux tube than in a wider flux tube at larger distances, where the magnetic field is stretched and weaker. The value of $B_{atm}$ is calculated by tracing the magnetic field line from the measurement positions in the magnetosphere to Jupiter's upper atmosphere, approximately $400~$km above Jupiter's dynamically flattened surface, as also described in \citeA{salveter2022}.

Since the best-fit parameters are prone to uncertainties, we calculate error-weighted means for the final energy flux results.
The error of the energy flux $\sigma_{F_E}$ is calculated, using the error propagation equation \cite[p.212]{taylor1997}, by
\begin{linenomath*}
\begin{equation}
\sigma_{F_E}^2 = \left(\frac{\partial F_{E}}{\partial A}\right)^2\sigma_A^2+ \left(\frac{\partial F_E}{\partial m}\right)^2\sigma_m^2+2\frac{\partial F_E}{\partial A}\frac{\partial F_E}{\partial m}\sigma_{mA}
\end{equation}
\label{eq:errorprop}
\end{linenomath*}
where $F_E$ is the equation for the total energy flux projected onto Jupiter's upper atmosphere, $\sigma_A$ and $\sigma_m$ are the errors of the fitting parameters, and $\sigma_{mA}$ is the covariance of the fitting parameters $m$ and $A$.
The weighted mean $\bar{F_E}$ is then calculated by
\begin{linenomath*}
\begin{equation}
\bar{F_E} =\frac{ \sum F_{E_i}/\sigma_{F_{E_i}}^2}{ \sum1/\sigma_{F_{E_i}}^2}
\end{equation}
\label{eq:weimean}
\end{linenomath*}
where $ F_{E_i}$ and  $\sigma_{F_{E_i}}$ are the total energy fluxes and the errors of all events considered.
The corresponding error in $\bar{F_E}$ is given by
\begin{linenomath*}
\begin{equation}
\sigma_{\bar{F_E}} = \frac{1}{ \sum1/\ \sigma_{F_{E_i}}^2}
\end{equation}
\label{eq:weimeanerror}
\end{linenomath*}
The definition of the weighted mean and its uncertainty are defined in \citeA[p. 177]{taylor1997}.

\subsection{Pitch Angle Diffusion} \label{pitch angle diffusion}
\begin{figure}[ht!]
\centering
\noindent\includegraphics[width=0.75\textwidth]{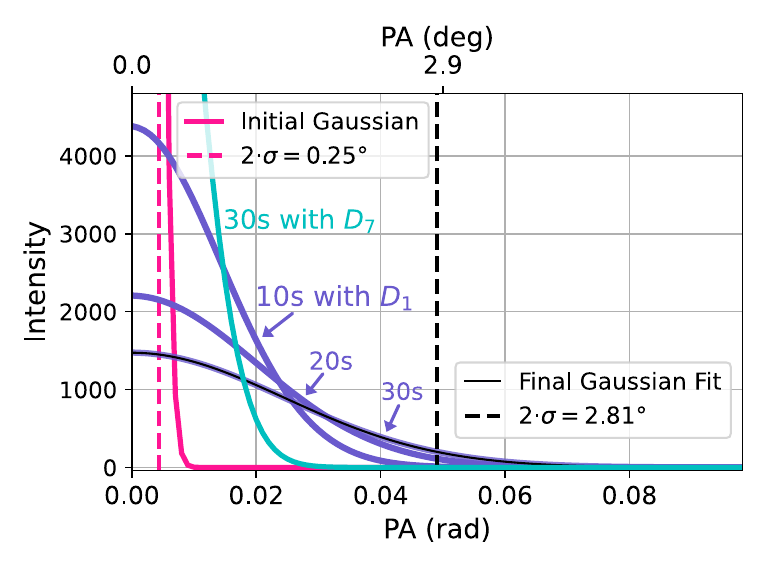}
 \caption{The pitch angle diffusion process is demonstrated for a diffusion coefficient $D_1 = 10^{-5}\mathrm{s}^{-1}$ from \citeA{li2021} for $\bar{E_1}=32~\mathrm{keV}$. The initial condition, a Gaussian peak with $2~\sigma$ corresponding to the mean loss cone in the middle magnetosphere, is shown in pink. The solution functions are shown in purple for three example times ($10 ~\mathrm{s}$, $20 ~\mathrm{s}$, $30 ~\mathrm{s}$), with the last solution fitted with a Gaussian function. The resulting $2~\sigma$ position at $\alpha= 2.81\degree$ is marked as dashed black line. In addition the solution for a higher energy channel diffusion coefficient $D_7 = 10^{-6}\mathrm{s}^{-1}$ for $\bar{E_7}=755~\mathrm{keV}$ is shown as turquoise line.}
\label{fig:diffusion}
\end{figure}

Electrons moving from the auroral altitudes to the equatorial region are assumed to be pitch angle scattered. To investigate its effect, we apply the pitch angle diffusion equation \cite{petschek1966}
\begin{linenomath*}
    \begin{equation}
        \frac{\partial f(\alpha,t)}{\partial t}=\frac{1}{\sin(\alpha )}\frac{\partial}{\partial \alpha }\left[ D \sin(\alpha)\frac{\partial f(\alpha,t)}{\partial \alpha}\right]
    \end{equation}
\end{linenomath*}
This diffusion equation is solved by applying the implicit backward Euler scheme for different sets of diffusion coefficients $D$. We assume as initial condition a Gaussian peak with a maximum value at $\alpha =0 \degree$ and a standard deviation of $2 ~\sigma =0.25\degree$, which corresponds to the mean local loss cone at the measurement positions. This characterizes the scattering process as a \add{simplified} diffusion process starting from the estimated width the beam could have in the middle magnetosphere without any pitch angle scattering. A Neumann boundary condition is used, which assumes zero change in electron intensity at the inner boundary $\partial f (0\degree)/\partial \alpha = 0$, along with a Dirichlet boundary condition, which assumes a constant electron intensity of zero at the outer boundary $f(90\degree) = 0$. The solution for time $t$ is fitted with a Gaussian function to determine the pitch angle position of the new $2~\sigma$ as the width of the scattered beam $\alpha_{max}$.
The diffusion process of the Gaussian beam signal is shown in Figure \ref{fig:diffusion} with the initial peak in pink and the solutions for three example times in purple. The last solution for $t = 30~\mathrm{s}$ is fitted and the resulting position of $2~\sigma$ is marked as a dashed black line at $\alpha= 2.81\degree$. The diffusion coefficients are taken from the work of \citeA{li2021}, which is described below.

The scattering time $t$ is calculated for each electron beam measurement using the traced propagation distance from the assumed acceleration region at a radial distance of $r = 1.2~\mathrm{R_J}$ to the middle magnetosphere position and the estimated velocity of the electrons. The velocity is simply estimated according to the relativistic kinetic energy equation \cite[p. 1137]{halliday} with the central energy of the energy channel considered.
%According to the findings of \citeA{salveter2022} the acceleration region is set to $r = 1.2~\mathrm{R_J}$.
The acceleration region is set to $r = 1.2~\mathrm{R_J}$ because the study of \citeA{salveter2022} is based on JEDI measurements at $r \geq1.2~\mathrm{R_J}$ and the measured upward electrons must have been accelerated below the spacecraft.

One set of diffusion coefficients $D_{\mathrm{Li}}$ is taken from the work of \citeA{li2021}, in which the pitch angle scattering of electrons driven by whistler mode waves is investigated as part of the generation of diffuse aurora.
The authors calculate the bounce-averaged pitch angle diffusion coefficients at the equatorial loss cone based on the measured whistler mode wave properties from the statistical spectra of \citeA{li2020}.
The scattering process modeled in \citeA{li2021} was found to be consistent with observations for electrons within the energy range investigated in our study.
The coefficients for each energy channel have been obtained from plot d) of Figure S3 in their supporting information for $M$-shells 17-20, which is the closest available position to our investigation region.
%....similar range to eelioot ?
In addition, the diffusion coefficients $D_{\mathrm{WM}}$ from \citeA{williams1997} are used for comparison, as these were obtained using a different methodology. 
The authors fitted electron pitch angle distributions with a solution of the pitch angle diffusion equation to determine the diffusion coefficients.
The focus of their study is the pitch angle scattering of electrons into the loss cone in the region above Ganymede.
Both sets of diffusion coefficients can be found in Table \ref{tab:diffusion} for each central energy.  
\citeA{elliott2018a} observed pitch angle scattering of electron beams, similar to the process considered in our study, in Jupiter's polar cap, the region poleward of the main aurora.
The diffusion coefficients in their study have values between both sets of diffusion coefficients. These coefficients were calculated independently of energy and are therefore not used in our study.
\add{For longer diffusion times the bounce averaged diffusion equation could be used for a more precise calculation} \cite{lyons1972}\add{, which considers the adiabatic pitch angle variation along the field line.}

The measurement positions used in this study are confined to the region surrounding the plasma sheet. These positions correspond to a range of magnetic latitudes that still exhibit variations in the magnetic field strength $B$. As an electron moves toward the magnetic equator, the decrease in $B$ results in a decrease in the pitch angle $\alpha$. 
This relation is described by
\begin{linenomath*}
\begin{equation}
\frac{\sin^2(\alpha_2)}{\sin^2(\alpha_1)}=\frac{B_2}{B_1}
\end{equation}
\end{linenomath*}
with magnetic field strengths $B_1$ and $B_2$ and pitch angles $\alpha_1$ and $\alpha_2$ of a particle at different locations  \cite[p. 25]{baumjohann}.
This effect could influence the beamness of studied electron beams if observations are considered at a larger range of magnetic latitudes, resulting in a change in pitch angle comparable to diffusion effects.
We demonstrate that for our observation region the estimated diffusion effects are more relevant to the change in $\alpha$ than the effects of the variation in $B$ at differing magnetic latitudes.
We consider $B$ at the largest magnetic latitude observed in our dataset and at the magnetic equator position traced along the field line from that position.
For an initial pitch angle of $0.25\degree$, the change in $B$ results in a decrease of $0.18\degree $. In contrast, the smallest diffusion coefficient $D_{\mathrm{Li}}=1\times 10^{-6}$ for $\bar{E_i}=755~\mathrm{keV}$ results in a mean increase in pitch angle of $0.6 \degree$.

\begin{table}
\caption{Diffusion coefficients $D_{\mathrm{Li}}$  from \citeA{li2021} and $D_{\mathrm{WM}}$  from \citeA{williams1997} for each energy channel central energy $\bar{E_i}$.}
\centering
\begin{tabular}{|ccc|}
\hline
 $\bar{E_i}$ (keV)&$D_{\mathrm{Li}}$ (s$^{-1}$)& $D_{\mathrm{WM}}$ (s$^{-1}$)\\
\hline
\hline
  $32$  & $1\times 10^{-5}$  & $5\times 10^{-4}$\\
  $57$  & $7\times 10^{-6}$  & $6\times 10^{-4}$\\
  $102$  & $5\times 10^{-6}$  & $4.9\times 10^{-4}$\\
  $180$  & $4\times 10^{-6}$  & $2.8\times 10^{-4}$\\
  $335$  & $3\times 10^{-6}$  & $1.5\times 10^{-4}$\\
  $755$  & $1\times 10^{-6}$  & $6.6\times 10^{-5}$\\
\hline
\end{tabular}
\label{tab:diffusion}
\end{table}

\section{Results}
\label{results}
In this section we present the results of our analysis. 
This includes a presentation of the beam detection method for an example measurement day and the resulting beam properties, such as the bidirectionality, the appearance duration of the beam phases, the distribution of beams as a function of the spatial parameters, 
the energy fluxes, and the relation between the intensity and the beamness parameter.

\subsection{Example Day of Measurement Properties}

\begin{figure}[h!]
\centering
\noindent\includegraphics[width=\textwidth]{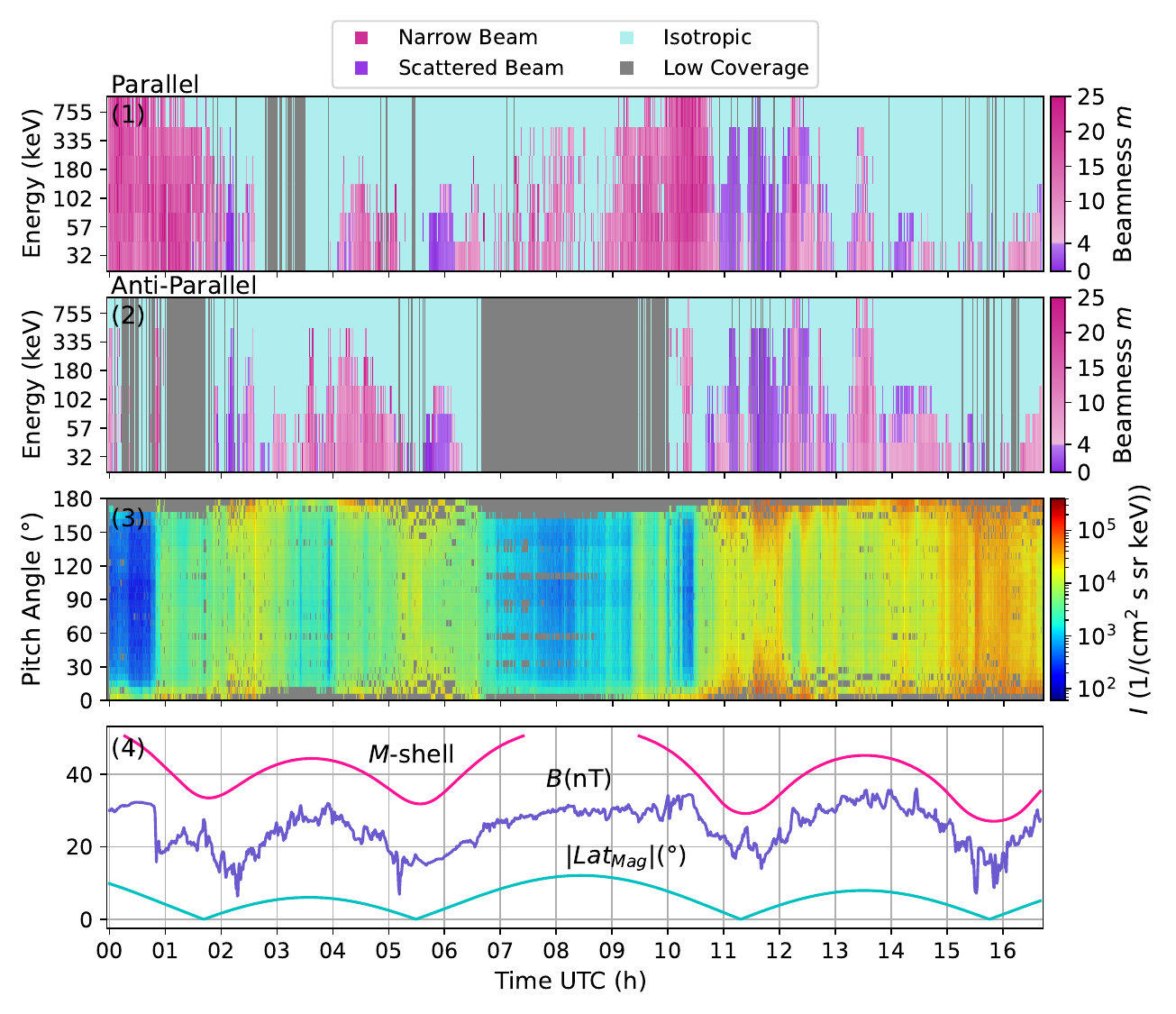}
\caption{Example overview plot for day 253 of 2019. (1) (2) Beam detection classifications of pitch angle distributions per energy channel and 30-s measurement interval for the parallel and anti-parallel directions, respectively. The beamness parameter is color coded, with narrow beams with $m\geq4 $ displayed in pink and scattered beams with $m<4$ displayed in purple. Intervals that did not meet the beam detection criteria, representing insufficient pitch angle coverage (Low Coverage) and low field-aligned intensity (Isotropic) are displayed in gray and blue, respectively. (3) The electron intensity-pitch angle spectrogram averaged over 30‐s intervals and over energies from $\sim30$ to 1,200 keV. (4) Measured total magnetic field strength $B$ in nT, the modeled $M$-shell in $\mathrm{R_J}$ and the absolute magnetic latitude in degrees.}
\label{fig:overview}
\end{figure}

Figure \ref{fig:overview} shows an overview plot of the example measurement day 253 of 2019 from 00 to 17 UTC that illustrates the functionality of the beam detection method and shows the properties of the measured environment.
The upper panels (1) and (2) show the fitting beamness parameter $m$ per 30-second measurement interval, as a function of the energy channel on the y-axis and time on the x-axis, for the parallel and anti-parallel directions, respectively. The beamness parameter is color coded, with narrow beams with $m\geq4 $ displayed in pink and scattered beams with $m<4$ displayed in purple. For this overview plot, the parameter error $\sigma_m$ is not considered. Measurement intervals where the field-aligned intensity is not sufficiently above the background electron intensity, as determined by the beam detection criterion two described in Section \ref{beam detection}, are denoted as 'Isotropic' and displayed in blue.
Measurement intervals with insufficient pitch angle coverage according to criterion one are denoted as 'Low Coverage' and displayed in gray.
Panel (3) shows the electron intensity-pitch angle spectrogram with intensity averaged over 30‐s intervals and over the six energy channels within $\sim30$ to ~$\sim1,200~ \mathrm{keV}$ as a function of time on the x-axis and pitch angle on the y-axis.
In addition to these electron characteristics, the final panel (4) shows magnetic field properties. This includes the measured total magnetic field strength $B$ in nT, the modeled $M$-shell in Jovian radii $\mathrm{R_J}$, and the absolute magnetic latitude based on the VIP4 model in degrees.

On this day electron beams are present throughout the whole day with some interruptions by isotropic distributions. The beams can also be seen visually in the spectrogram, indicated by higher intensities near $\alpha=0\degree$ and $\alpha =180\degree$ than for pitch angles around $90\degree$, particularly between 10 and 17 UTC.
The high temporal resolution of the JEDI measurement allows for the conclusion that beam phases occur on timescales of a few minutes to a few hours, interrupted by isotropic distributions, similar to what was reported by \citeA{mauk2007} based on Galileo electron data. A more detailed study of the duration of beam phases is presented in Section \ref{durationofbeamphases}.

It is evident that the majority of the beams are detected in the lower energy channels. Moreover, as the energy increases, the number of beams decreases. This reveals a pattern showing that beams in the highest energy channel are usually present only when beams are also detected in all lower energy channels. This can be attributed to the lower detectability of higher energy beams due to their smaller intensity. One possible physical reason is that beams with higher energy experience less scattering, as indicated by the decreasing pitch angle diffusion coefficients with increasing energy in Table \ref{tab:diffusion}. Consequently, these beams may tend to be more narrow in general.
This implies that their detectability depends more on sufficient pitch angle coverage close to the field-aligned direction compared to stronger scattered, broader, low-energy beams. 
Furthermore, a smaller degree of scattering would result in a larger portion of the beam electrons overcoming the mirror point and getting lost in Jupiter's atmosphere, leading to a decline in the number of trapped higher energy beams.

The periodic features in the intensity spectrogram and the magnetic field properties are clearly related to the spin of Jupiter, which has a period of around 10 hours, and can be attributed to the motion of Juno relative to the magnetic equator, as illustrated in Figure \ref{fig:orbits}. 
In the vicinity of the magnetic equator, the magnetic field strength $B$ reaches local minimum values.  
The background electron intensity, characterized by intensities around $\alpha = 90 \degree$, increases with decreasing distance to the magnetic equator, indicating an approach or crossing of the plasma sheet.
Both observations were previously documented in a statistical study of the plasma sheet by \citeA{liu2021}. %(maybe also MA 2021?)
The $B$ minima show a correlation with the magnetic latitude and the $M$-shell values. At the modeled magnetic equator the $M$-shell is equal to the radial distance and therefore shows local minima for these times. This correlation indicates that the $M$-shells are sufficiently modeled.
As mentioned by \citeA{ma2021} the beam distributions are more scattered close to the plasma sheet due to the increased intensity of trapped background electrons. This can be observed in our study by an increase in the number of purple intervals close to the $B$ minima, showing an increased presence of scattered beams with $m<4$.

\subsection{Bidirectionality of beams}
The example day demonstrates that the pitch angle coverage criterion is often met in only one direction, parallel or anti-parallel, depending on the local instrument configuration relative to the magnetic field. 
If both directions show sufficient coverage, beams are often detected in both directions, showing bidirectional beams rather than monodirectional beams.
It is discussed that the auroras observed on Jupiter's north and south poles might be conjugated, meaning that auroral features are simultaneously generated in both hemispheres \shortcite{gerard2013}. If this is the case, then it would suggest that the acceleration of electrons occurs approximately in parallel in the northern and southern auroral acceleration regions.
Consequently, the majority of auroral beams detected in the middle magnetosphere would be bidirectional, as is indeed reported from previous observations.
We performed a more detailed study of the bidirectionality of beams, including all measurement intervals used. The results of this study are shown in Table \ref{tab:monobi}.
The first column shows the central energy of the energy channels $E_2$ to $E_7$. Columns 2 and 3 show the relative amount of bidirectional and monodirectional narrow beams for each energy channel $E_i$, respectively. We only consider narrow beams that satisfy the narrow beam definition with the error of the beamness parameter ($m - \sigma_m \ge 4$), as described in Section \ref{beam detection}. The occurrence rates are calculated relative to the total number of intervals, satisfying the pitch angle criterion in both directions, ensuring that a bidirectional beam could have been measured. For the monodirectional occurrence rate, the counts of parallel and anti-parallel beams are added.  

For energy channels $E_2-E_6$, more bidirectional than monodirectional narrow beams have been detected. However, in the highest channel $E_7$ the occurrence rate of monodirectional beams exceeds that of bidirectional beams.
Note that the number of detected beams decreases significantly with increasing energy, perhaps as a result of the decreasing intensity of the beams with increasing energy, resulting in decreased detectability. Consequently, the statistical distribution is less representative for the higher-energy channel beams. This could explain the contrasting case of more monodirectional beams for channel $E_7$.

Intervals classified as monodirectional beams exhibit better pitch angle coverage on average in the direction of the detected beam compared to the isotropic direction.
The mean smallest pitch angle  observed with respect to the magnetic field line is $10\degree$ in the directions classified as potentially isotropic. This value is significantly higher than the mean smallest pitch angle observed in the directions of the detected beams, which is $5.2\degree$. These findings suggest that some pitch angle distributions may be classified as monodirectional beams due to asymmetric field-aligned pitch angle coverage, that is, the coverage was insufficient in one direction to detect a narrow beam.
Therefore, we determine how many of the distributions showing monodirectional narrow beams exhibit symmetric pitch angle coverage in both directions.
We calculate the relative amount of monodirectional narrow beams where the smallest measured pitch angle in the non-beam direction is equal to or smaller than the smallest pitch angle covered in the beam direction. This ensures equivalent or similar detectability in both directions. These occurrence rates are shown in column 4 of Table \ref{tab:monobi}. Less than $\sim17\%$  of the previously classified monodirectional beams meet this criterion, indicating that the observed monodirectionality of beams may often be attributed to technical reasons. Whether all of these distributions are true monodirectional beams and what characteristics they exhibit must be studied in future work.
%This finding indicates the potential underestimation of detected beams due to insufficient pitch angle coverage.

\begin{table}[h!]
    \centering
\caption{Occurrence rates of bidirectional (column 2) and monodirectional (column 3) narrow beams ($m - \sigma_m \ge 4$) relative to the total number of intervals, satisfying the pitch angle criterion in both directions for each energy channel with central energy $\bar{E_i}$ (column 1). Column 4 shows the relative amount of monodirectional narrow beams for intervals in which the smallest measured pitch angle in the non-beam direction is equal to or smaller than the smallest pitch angle covered in the beam direction.}
\label{tab:monobi}
    \begin{tabular}{lccc}
    \hline
    $\bar{E_i}$ (keV)& Bidirectional~$(\%)$& Monodirectional~$(\%)$& \begin{tabular}{c} Monodirectional\\(similar PA)~$(\%)$\end{tabular}\\
    \hline
    
        32& 9.9&  6.0&  1.0\\
        57& 8.8&  6.6&  0.9\\
        102& 7.2&  6.4&  0.9\\
        180& 5.8&  5.4&  0.7\\
        335& 5.3&  4.5&  0.6\\
        755& 1.6&  3.1&  0.5\\
    \hline
    \end{tabular}
\end{table}
\subsection{Duration of Beam Phases} \label{durationofbeamphases}
The distribution of the beam detection classifications as a function of time in Figure \ref{fig:overview} demonstrates that the occurrence of beams is structured as described by \citeA{mauk2007}. The beams often occur in phases of minutes to hours interrupted by phases of isotropic distributions. We performed a more detailed analysis on the duration of the beam phases in order to determine the intermittency of the beams. The structuring of the beam phases can provide insight into the persistence and structure of the acceleration mechanism close to the planet, given the hypothesis that the beams originate from the auroral acceleration region. 

It is important to note that the duration of detected beam phases is limited to the duration of phases with sufficient pitch angle coverage. 
For the analysis, we define phases of good pitch angle coverage as a minimum of three measurement intervals in sequence (90 s) with sufficient pitch angle coverage according to the first beam detection criterion. For each of these phases, beam phases and isotropic phases are determined. We define a beam phase as one or multiple measurement intervals in sequence for which a beam, scattered or narrow, is measured in at least one of the energy channels. Since we want to analyze how intermitted the beam phases are, narrow and scattered beams are considered together.
Otherwise, shorter duration times would be calculated for a narrow beam phase in which scattered beams are present, as in the vicinity of the plasma sheet.
We define an isotropic phase as one or more measurement intervals in sequence for which isotropic distributions are detected in all energy channels.
It is evident that beam phases, which are as long as the phase of sufficient pitch angle coverage, might be longer than calculated. 
How much longer the beam phase might be can not be determined. Nevertheless, phases that have definitively ended can be identified.
Therefore, we introduce the real beam phase as an additional phase classification. This phase is characterized by its true length, with its beginning and end observed within a longer phase of sufficient pitch angle coverage.
This means that the detection of at least one measurement interval without beams is required before and after the beam phase.
According to this definition, phases that end due to insufficient pitch angle coverage are excluded in this classification.

\begin{figure}[ht]

\centering
\noindent\includegraphics[width=0.85\textwidth]{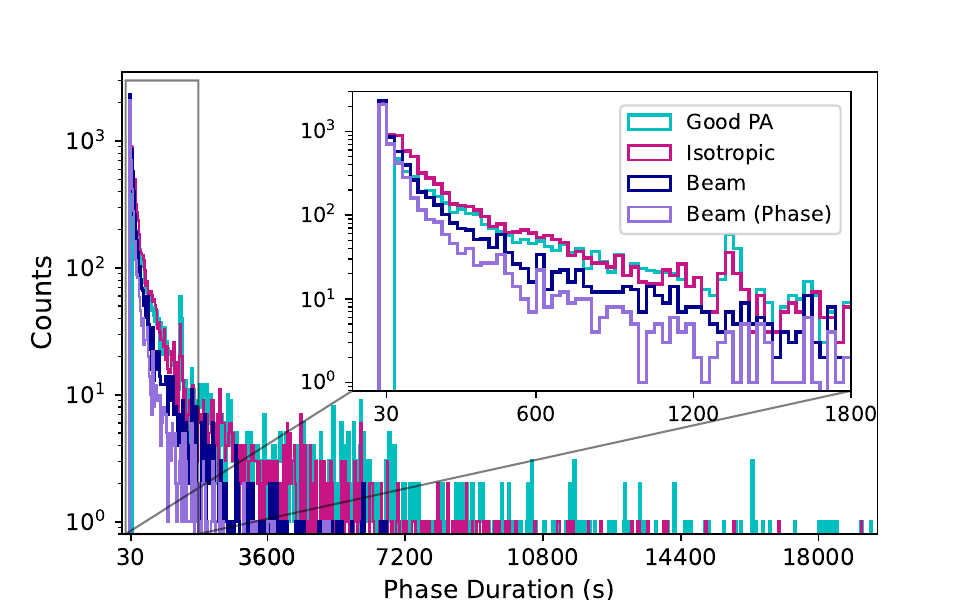}
\caption{Histogram of the four types of phase with phase counts as a function of the phase duration in seconds. The phase with sufficient pitch angle coverage is labeled as 'Good PA', the isotropic phase as 'isotropic', the beam phase with 'Beam' and the real beam phase as 'Beam (Phase)'.}
\label{fig:duration}
\end{figure}

Figure \ref{fig:duration} shows a histogram of the four types of phases with phase counts as a function of the phase duration in seconds. The phase with sufficient pitch angle coverage is labeled as 'Good PA', the isotropic phase as 'Isotropic', the beam phase with 'Beam' and the real beam phase as 'Beam (Phase)'.
The maximum phase length is 5~h and 23~min (19380~s) for the 'Good PA' phases, 5~h and 19~min (19140~s) for the isotropic phases, 1~h and 52~min (6720~s) for the beam phases and 1~h and 27~min (5220~s) for the real beam phase.
All phase types show decreasing counts with increasing phase duration. The good pitch angle phase and the isotropic phase show in general the highest counts, followed by the beam phases and real beam phases.
The high amount of short beam phases may indicate that the auroral electron acceleration toward the magnetosphere, and consequently also toward Jupiter, is not a continuous process but is structured, as suggested by \citeA{mauk2007}. As demonstrated in Figure \ref{fig:overview}, isotropic distributions are detected throughout the day, even during continuous beam phases, at least in the highest energy channel. Some beam phases may be interrupted by short isotropic intervals, which include unresolved beams, resulting in a higher number of shorter beam phases. Nevertheless, the existence of multiple isotropic phases with long phase durations in the middle magnetosphere provides further support for the concept of a structured auroral acceleration process.

\subsection{Spatial Beam Distribution} \label{spatial beam distribution}
\begin{figure}[h!]
\centering
\noindent\includegraphics[width=1\textwidth]{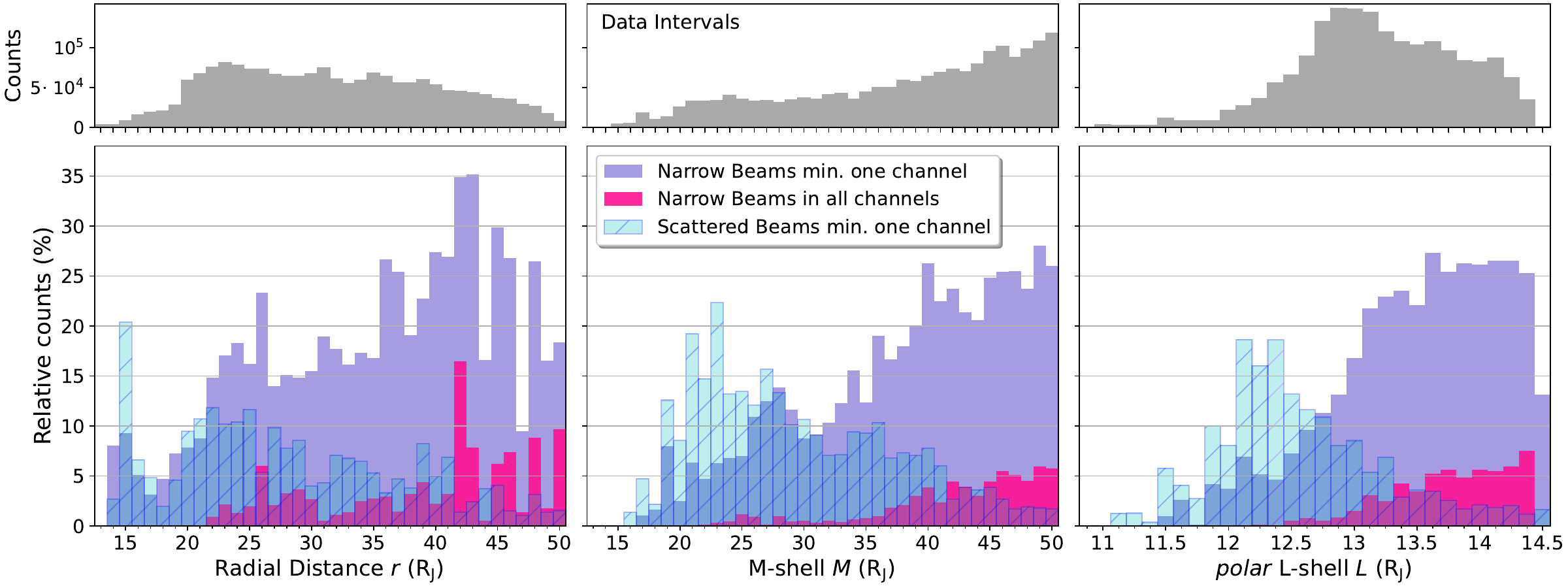}
\caption{Histogram of pitch angle distribution classifications as function of radial distance $r$ ($\mathrm{R_J}$), $M$-shell and the $polar~L$-shell. The upper plots show the absolute counts of data intervals that met the pitch angle coverage criterion (gray).
The lower three plots show the distribution of different pitch angle distribution classifications. Counts of measurement intervals with narrow beams detected in a minimum of one energy channel (violet), narrow beams detected in all energy channels (pink), and scattered beams in a minimum of one energy channel (blue) are displayed relative to the counts of intervals with sufficient pitch angle coverage. }
\label{fig:beamrad}
\end{figure}

The statistical occurrence of pitch angle distribution classifications is presented in Figure \ref{fig:beamrad} as a function of the radial distance $r$, the $M$-shell and the $polar~L$-shell.
The absolute counts of data intervals that met the pitch angle coverage criterion are shown in the upper three plots as gray bars.
The lower three plots show the distribution of different classifications of pitch angle distributions.
The violet bars represent the relative counts of measurement intervals for which a true narrow beam with $m - \sigma_m \ge 4$ is detected in at least one of the energy channels. The pink bars represent the relative counts of intervals for which narrow beams are detected in all energy channels.
The dashed blue bars represent the relative counts of measurement intervals for which a scattered beam with $m + \sigma_m < 4$ is detected in at least one energy channel.
The classification amounts in each bin are calculated relative to the number of measurement intervals that met the pitch angle coverage criterion in this bin.
The relative counts are calculated considering both the parallel and anti-parallel directions, as the occurrence distributions of both directions shows a high degree of similarity resulting from the bidirectionality of the beams.

A total of $180,543$ intervals with sufficient pitch angle coverage, $33,422$ intervals with narrow beams in a minimum of one channel, and $5,180$ intervals with narrow beams in all channels are identified. 
The measured radial distances correspond to a similar range of $M$ values from 14 to 50, with a bin width of $1~\mathrm{R_J}$ and to $polar~L$ values ranging from 11 to \change{14.75}{14.5}, with a bin width of $0.125~\mathrm{R_J}$.
The $M$-shell parameter describes the more realistic and strongly stretched magnetic field lines in the middle magnetosphere. Consequently, they correspond to a much smaller range of $polar L$-shells, which describe a dipole magnetic field. As discussed in Section \ref{spatial parameters}, the $polar ~L$-shell calculation is performed to determine the observation range of the auroral studies of \citeA{salveter2022} that magnetically maps to the measurements used in our study. It is not expected to accurately reflect the $polar~L$-shell. However, a notable similarity is evident in the course of the $polar~L$ and $M$ distributions.

The distribution as a function of the radial distance shows that narrow beams are present throughout the observation range between 14 and 50 $\mathrm{R_J}$. In the distribution as a function of $M$-shell, narrow beams are detected for $17\leq M\leq50$.
A clear trend is evident in all distributions, characterized by an increase in the relative number of narrow beams with increasing distance. While the amount decreases from 43 $\mathrm{R_J}$ in the radial distance distribution, it reaches a maximum close to the observation edge for the $M$-shell distribution. 
The observed trend is more pronounced and shows a smoother curve with fewer outliers for the $M$ distribution than for the radial distance.
Hence, the $M$ distribution appears to represent more physically coherent trends.
The $M$-shell and $L$-shell values are derived from model calculations which include uncertainties and do not represent a true positional measurement such as the radial distance distribution. 
However, it can be assumed that the $M$ distribution is more reliable, given that the magnetic field line tracing is performed over distances restricted to the equatorial region and the distribution exhibits a high degree of similarity with the true radial distance distribution.
The narrow beams detected in all energy channels show comparable trends, with an increase in relative counts with increasing distances, but without counts for $M <23$.
An inverse trend is evident in the scattered beams, characterized by a general decrease in the relative counts for increasing distances for $M\geq23$. Around $M=23$, the scattered beams exhibit a pronounced maximum.
It is important to note that with a limited number of total data intervals used, for example around $M=15$, the calculated distributions may not accurately reflect the real beam distribution.

The primary focus of this study is on the true narrow beams with a beamness $m$ that satisfies the condition $m - \sigma_m \ge 4$.
These beams are hypothesized to be auroral electron beams, as their narrow width indicates their generation close to the planet \cite{saur2006}. Therefore, they are assumed to represent a statistical distribution comparable to that of auroral upward electrons.
The scattered beams with $m+\sigma_m<4$ may be auroral beams, which have undergone stronger pitch angle scattering as they propagated through the magnetosphere.
As described in \citeA{frank2002}, it is assumed that over the lifetime of a beam its pitch angle distribution changes due to scattering processes from a narrow beam to a more scattered beam, until it shows isotropic distributed intensity. Such a scattering process can occur during the propagation of the beam through the magnetosphere if the electrons remain outside the loss cone and bounce back and forth between the two hemispheres.
In the absence of pitch angle scattering, the beam electrons would remain at small pitch angles within the loss cone and reach Jupiter's atmosphere at the opposite hemisphere.
However, the gradual increase in intensity from $\alpha \approx90\degree$ to field-aligned pitch angles of the scattered beam distributions could be the signature of a distinct generation, for example by adiabatic heating processes \cite{woodfield2014,nenon2022}.

The spatial distributions presented demonstrate that electron beams are present throughout the observation range in the middle magnetosphere, consistent with previous reports \cite{tomas2004a,tomas2004b,frank2002,mauk2007,ma2021}. The decreasing amount of narrow beams at smaller distances is consistent with the gradual transition of bidirectional or isotropic to pancake pitch angle distributions that has been previously discussed for $M<15$ \cite{tomas2004a,ma2021}.
% connection to aurora
The main auroral emission can be mapped to the equatorial magnetosphere between the orbit of Ganymede at $r \approx 15 ~\mathrm{R_J}$ and several tens of Jovian radii, solely based on the position of Ganymede's auroral footprint equator-ward of the aurora \cite{tomas2004a}.
As discussed by \citeA{allegrini2020}, the main auroral emission is generally assumed to map to $20-60 ~\mathrm{R_J}$, but the mapping depends on the chosen magnetic field model. These authors find the equatorward edge of the main auroral emission to map to $M$-shells of 51 using the JRM09 model \cite{connerney2018} and the CAN1981 model \cite{connerney1981}. However, they also state that an improved magnetic field model will likely bring this boundary closer to the expected $20-30~\mathrm{R_J}$.
Therefore, the detection region of the narrow beams described in this study, as well as their narrow angular width, already indicates that their source may be located in auroral acceleration regions close to Jupiter, as previously postulated \cite{frank2002,mauk2007}. 
% vergleich mit annika

For a comparison between the occurrence distribution of auroral electrons and the middle magnetosphere beams, we focus on the detected beams in a minimum of one energy channel.
This is because the detection method has on numerous occasions classified pitch angle distributions, especially those in the higher-energy channels, as isotropic, which could include narrow, low intensity beams, which are not resolved. This may be a consequence of the field-aligned intensity criterion that was chosen in order to exclude distributions with random high field-aligned intensity. 
A comparison of the $L$-shell distribution with the results shown in Figure 7 of \citeA{salveter2022} reveals that the electrons investigated in our study magnetically map to the smaller part of the polar observation range between $L$-shells of $11-15$. 
The measurements cover only a small range of $polar~L$-shells and cannot be definitively associated with the specified position, as discussed in Section \ref{spatial parameters}. A similar $L$-shell range is determined to correspond to $M$-shells between 14 and 50 in the supporting information of \citeA{salveter2022}.
Furthermore, the relative counts of the field-aligned distributions (Figure 5) and relative to them the upward and bidirectional electrons (Figure 7) in \citeA{salveter2022} increase consistently and significantly with distance in the covered $L$-shell range and beyond. This trend corresponds to a recognizable feature, which is indeed identified in our beam distributions, which is consistent with the hypothesis tested that the middle magnetosphere beams represent upward accelerated auroral electrons.
It is possible that the true increase in the relative number of beams with distance may be even more pronounced, given that electron distributions generally exhibit a decrease in intensity and consequently detectability with increasing distance from the planet.

The inverse trend of scattered beams compared to narrow beams seems surprising at first.
An increasing number of narrow beams with increasing distance could be expected to result in an increasing number of scattered beams under the assumption that narrow beams transform into scattered beams.
%Possible explanations for this result are discussed below.
The observed decrease may be attributed to the technical aspect that scattered beams are sorted out faster by the beam detection method than narrow beams.
This is because the field-aligned intensity is compared to the mean isotropic background intensity in a wide pitch angle range from $45\degree -135\degree$ that was adapted for the detection of narrow beams. However, for strongly scattered beams, the specified pitch angle range contains the beam itself. Therefore, the mean intensity is larger, and a scattered beam must show a larger field-aligned intensity than a narrow beam to satisfy the second criterion.
In addition, the detectability of beams decreases with distance as their intensity decreases. 
%It is evident that both factors affect the calculated distribution of scattered beams.
However, it remains unclear to what extent this effect increases with distance and if it would cause a reverse trend in an actual increasing distribution.
In contrast, the maximum of the scattered beams around $M=23$ is particularly pronounced and is also evident in the results of \citeA{ma2021}, suggesting that it is less attributable to technical factors.
In their study the pitch angle distributions undergo a gradual transition from field-aligned to pancake-like, with broader field-aligned distributions in between at $M\approx20$.

As previously described in this section, the broader beams may not only show scattered auroral beams but also originate from plasma transport mechanisms.
Plasma transport in Jupiter's magnetosphere, from the plasma source region surrounding Io's orbit to the outer magnetosphere, is described in terms of the centrifugally driven flux tube interchange \cite{southwood1987}.
In the context of electron transport in Saturn's magnetosphere, \citeA{rymer2008} state that, with the conservation of the first two adiabatic invariants, outward radial plasma transport of an isotropic pitch angle distribution leads to field-aligned distributions, similar to beams with $m\approx0$, as the electrons cool. In contrast, inward radial transport results in a heated pancake distribution.
\citeA{nenon2022} discuss that the outward adiabatic transport of initially energetic electrons has the potential to generate energetic scattered beam distributions.
As demonstrated by \citeA{woodfield2014}, the acceleration of electrons via whistler waves can generate such a source of energetic electrons. This results in a phase space density peak around $L\approx9$ and drives outward radial diffusion of electrons for larger $L$-shells.

The plasma transport processes may also provide an explanation for the significant presence of broad field-aligned beams in alignment with the assumption that these distributions are pitch angle scattered auroral beams.
\citeA{clark2014} showed for Saturn's magnetosphere that the inward adiabatic transport of field-aligned electron distributions has the capacity to broaden a bidirectional beam towards a butterfly distribution, ultimately resulting in a pancake distribution.
As demonstrated in \citeA{kollmann2018}, the inward radial diffusion of energetic electrons may be a significant mechanism within the middle magnetosphere, responsible for populating the Jupiter radiation belts with highly energetic electrons. As suggested by these authors, a source of energetic electrons is required at larger distances, which could be explained by the acceleration of electrons in the auroral acceleration regions.

Another possible reason is based on the timescales of plasma transport.
As demonstrated by \citeA{bagenaldelamere2011}, plasma transport is slowest deep in the magnetosphere near Io and increases with distance.
Beams that are generated at smaller $M$-shells will thus remain close to these $M$-shells for a longer duration, and can be subject to a greater degree of pitch angle scattering until they are transported out of the middle magnetosphere than beams that are generated at larger $M$-shells. 
The radial plasma transport speed is calculated as $6-30 ~\mathrm{km~s^{-1}}$ and $20-150 ~\mathrm{km~s^{-1}}$ at radial distances of $r=20~\mathrm{R_J}$ and $r=40~\mathrm{R_J}$, respectively \cite{bagenaldelamere2011}.  We estimate the radial distance that an initial narrow beam would be transported outward until it is scattered to a width considered as the width of the scattered beam ($m<4$) experiencing a continuous pitch angle scattering.
The required scattering time is derived for a more simplified diffusion process, starting from an initial distribution with a one-half width at the half maximum value (HWHM) of $0.2\degree$ to a distribution with a HWHM of $9.9\degree$, which is the HWHM of the beam function for the threshold beamness of $m=4$ \cite{mauk2007}. The calculation is performed representatively using the diffusion coefficients $D_{\mathrm{Li}}=5\times10^{-6}$ and $D_{\mathrm{WM}}=4.9\times10^{-4}$ of Table \ref{tab:diffusion} for electrons with a kinetic energy of $102 ~\mathrm{keV}$. This yields diffusion times of $t_{Li}=5732$~s and $t_{WM}=58$~s, compared to averaged bounce periods of about 90~s. The resulting radial distance traversed by a beam as it moves with the flux tube for these times ranges from $0.5 - 2.4 ~\mathrm{R_J}$ and  $1.6 - 12 ~\mathrm{R_J}$ for an initial beam generation at $M=20$ and $M=40$ , respectively, considering $D_{\mathrm{Li}}$. For $D_{\mathrm{WM}}$ the radial distance ranges from $0.005 - 0.024 ~\mathrm{R_J}$ and $0.016 - 0.12 ~\mathrm{R_J}$ for an initial beam generation at $M=20$ and $M=40$, respectively. 
The calculations demonstrate that, for the faster transportation scenario and diffusion coefficients similar to $D_{\mathrm{WM}}$, the closer beam would be transported to $M=22$, while the beam generated at a more distant location would be at $M=62$ and, consequently, would fall outside the range of observation. This estimation suggests that the beams located in the outer middle magnetosphere may not contribute significantly to the scattered beam population within our observation range. In the case of the slower scenario and diffusion coefficients closer to $D_{\mathrm{Li}}$ the contribution of scattered beams would be similar at $M=20$ and $M=40$.
These calculations are simplified and intend to demonstrate a possible influence of the increasing plasma transport velocity on the observed beam population in the middle magnetosphere. They do not consider additional effects, such as different transport conditions for higher energy electrons or a localized flux tube interchange at low latitudes instead of the interchange of two whole flux tubes \cite{ma2019}, which would only transport the low altitude portion of the beam electrons.

In this study, we cannot distinguish between a broad field-aligned distribution generated by adiabatic heating and a strongly scattered beam distribution that actually shows a pitch angle scattered auroral beam, since both appear to show pitch angle distributions with gradually increasing intensities towards field-aligned pitch angles.
Consequently, the underlying mechanism behind the formation of the broad beams cannot be clarified here. However, the narrow beams observed in our study are consistent with originating from the auroral acceleration region.

\subsection{Energy Flux}

\begin{figure}[h!]
\centering
\noindent\includegraphics[width=0.75\textwidth]{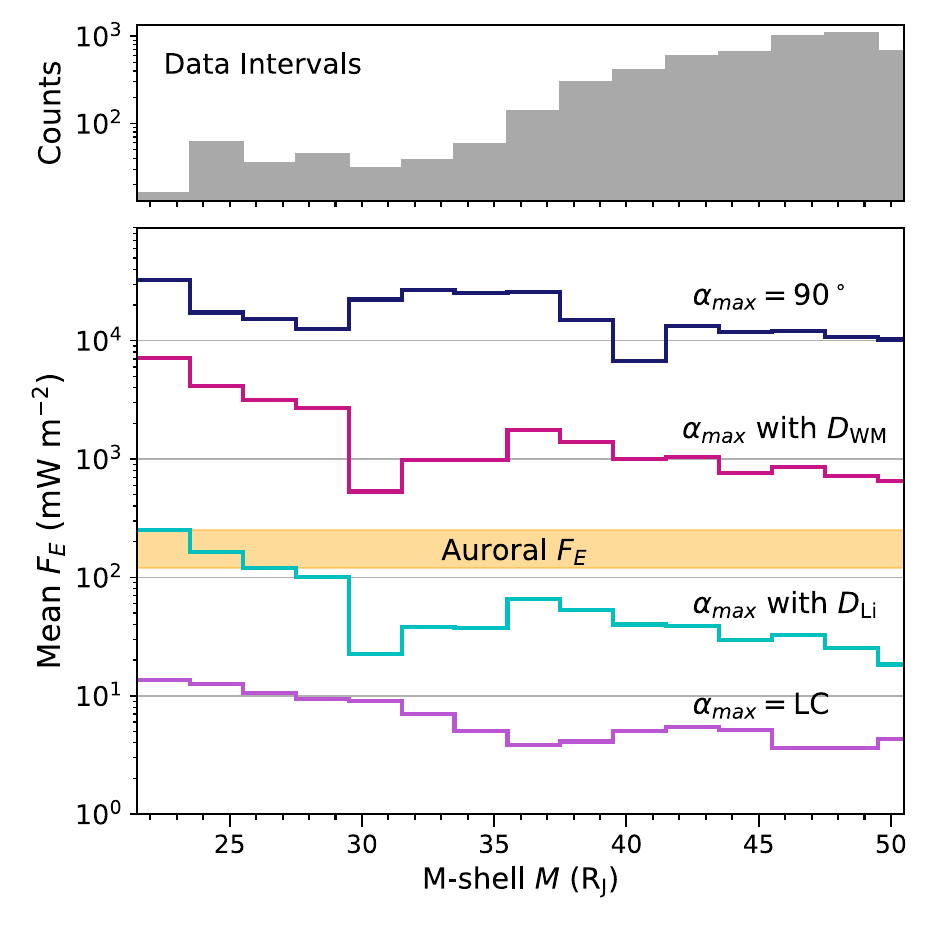}
\caption{Error weighted mean energy fluxes of the narrow electron beams projected into Jupiter's upper atmosphere as a function of $M$-shell. The upper plot shows the absolute counts of data intervals used for the energy flux calculations. We calculated the energy fluxes for four different beam width definitions $\alpha_{max}$, depending on the assumed pitch angle scattering process during the beam's propagation to the middle magnetosphere. The error weighted mean is calculated for each $M$-shell bin including both beaming directions. The orange shaded area shows the range of the mean energy flux of the auroral field-aligned electrons from \citeA{salveter2022} for the covered $L$-shells.}
\label{energyflux}
\end{figure}

In addition to the statistical occurrence of electron beams, their energetic properties can be employed to compare different electron distributions.
The lower plot in Figure \ref{energyflux} shows the energy fluxes of the beams projected into Jupiter's upper atmosphere for each of the four definitions for $\alpha_{max}$, which are described in Section \ref{calculation of the energy  flux}.
The upper plot shows the absolute counts of the data intervals used for the energy flux calculations, which correspond to the intervals in which true narrow beams are detected in all energy channels.
It is important to note that the energy flux is calculated using the fitting intensity function and not the measured intensities. This is necessary in the equatorial middle magnetosphere, where narrow beams are not perfectly resolved by JEDI, but this also leads to additional potential for uncertainties in the estimated intensity.
For each $M$-shell bin, the error weighted mean of the energy flux is calculated considering the error of each energy flux value as described in Section \ref{energyflux}.

The total beam energy flux with an integration limit of $\alpha_{max}=90\degree$ serves as an upper limit with values ranging from $ 7\times10^3$ to $3\times10^4~ \mathrm{mW~m^{-2}}$.
The lower limit, on the other hand, is determined by the energy flux considered within the local loss cone, exhibiting a mean value of $0.25\degree$. This estimates the width of the electron beam in the absence of any pitch angle scattering. The energy flux values are more than three orders of magnitude smaller, ranging from $4$ to $14 ~\mathrm{mW~m^{-2}}$.
As additional limits we calculated the scattered width of a new auroral beam experiencing pitch angle scattering during the $\sim1/4$ bounce period from the auroral region to the measuring position in the equatorial middle magnetosphere. Using the diffusion coefficients $D_{\mathrm{Li}}$  of \citeA{li2021}, the scattering width $\alpha_{max}$ ranges from a mean of $2.8\degree$ for the $32~\mathrm{keV}$ energy channel to $0.6\degree$ for the $755~\mathrm{keV}$ energy channel, resulting in energy fluxes between $2\times10^1 $ and $2.5\times10^2 ~\mathrm{mW~m^{-2}}$.
The energy fluxes within the scattering widths calculated with the comparison diffusion coefficients $D_{\mathrm{WM}}$  from \citeA{williams1997}, which range from a mean of $20\degree$ to $4.5\degree$, are closer to the total beam energy flux with values of $6.5\times10^2 $ and $7\times10^3 ~\mathrm{mW~m^{-2}}$.
For all four approaches, the projected energy flux generally decreases with increasing distance.
This may be an actual feature of the middle magnetosphere beams or it may result from the decreasing intensity caused by the decrease in magnetic field strength over distance, which may not be perfectly calculated out of the data using the magnetic flux conservation approach.
It must be considered that the absolute counts of intervals are low for $M<35$ and that the energy fluxes may be less representative. For this $M$-shell range, the energy flux curves are less smooth and generally show larger values, which may be characteristics of specific beams, rather than a statistically representative property.

Assuming that the detected middle magnetosphere beams represent the auroral upward field-aligned electrons, it is expected that they exhibit similar energy fluxes.
The energy flux in Figure 5 of \citeA{salveter2022} is calculated as the mean of all auroral field-aligned distributions, i.e. the upward and bidirectional accelerated electrons, which are probably represented by the detected beams in the middle magnetosphere, as well as the downward accelerated electrons.
However, the relative counts for the upward and bidirectional classifications, and the downward classifications are comparably around $50\%$ for the covered $L$-shells.
Consequently, the calculated mean energy flux values can be considered representative for the upward and bidirectional electrons and can be meaningfully compared with the energy fluxes of the beams in our study.
The mean energy flux of the auroral electrons for the covered $L$-shells ranges from around $1\times10^2$ to $3\times10^2 ~\mathrm{mW~m^{-2}}$. This range is included in Figure \ref{energyflux} as an orange shaded area.
In contrast to the occurrence distribution, the smaller L-shells do not exhibit a pronounced monotonic trend. Therefore, it is not expected that the course will be identified in our results, since the magnetic mapping serves only as a first approximation.
In our study, the energy flux within the local loss cone was found to be more than one magnitude smaller. This indicates that the majority of the beam electrons are scattered out of the loss cone, under the assumption that the intensities within the loss cone are accurately modeled.
Consequently, these electrons will not reach Jupiter's atmosphere. Instead, they will be reflected back and stay trapped in the magnetosphere, contributing to the population of Jupiter's magnetosphere with energetic electrons. 
\add{It is worth mentioning that a minor portion of electrons remaining in the loss cone  might be backscattered to the magnetosphere via coulomb scattering, as observed for electrons precipitating into Earth's atmosphere} \cite{qin2024}.
The total beam energy flux, calculated with $\alpha_{max}=90\degree$, is about two orders of magnitude larger than the auroral mean.
One potential explanation for this is that, even with a beamness parameter larger than four, considering only narrow beams, which are expected to be relatively newer, not only newly generated beams are measured, but a superposition of new beams and trapped parts of older beams.
Conversely, within the auroral acceleration region, observations are likely to measure only newly generated upward beams, since only the intensities in the loss cone are considered. 

In addition, technical reasons influence the value of the calculated energy flux in both studies.
To avoid an underestimation of the intensity, which is often not well resolved by JEDI at small pitch angles, we calculate the intensity from the fitting beam intensity function.
This is not necessary for the polar observations as the loss cones are much larger and better covered by JEDI measurements. 
Nevertheless, it is plausible that not the entire intensity is observed in the polar observation. The underestimation of the intensities of the electron beams is described in the supporting information of \citeA{mauk2018}. Consequently, the polar energy fluxes may be smaller than the actual mean energy flux.
Furthermore, as previously discussed, it can be assumed that in our study beams with lower intensities are more likely to be sorted out by the beam detection method, possibly resulting in a larger mean energy flux. 

The auroral mean energy flux falls between the energy fluxes within the scattering widths calculated with $D_{\mathrm{Li}}$ and $D_{\mathrm{WM}}$, the energy flux for $D_{\mathrm{Li}}$ being slightly closer to the auroral energy flux.
The similar energy fluxes seem to be consistent with the picture that the middle magnetosphere beams are generated in the auroral acceleration regions. 
In addition, they indicate that a numerical solution of the diffusion equation, using diffusion coefficients in between both sets of coefficients, yields a good first estimation of the scattering process of auroral beams. 
\add{It is important to note that the diffusion coefficients used are attributed to a pitch angle scattering processes based on wave-particle interaction. However, other mechanisms may also be relevant, such as field line curvature scattering, which can be significant for energetic electrons in the middle magnetosphere} \cite{gu2024}.
\change{It must be noted that,}{Furthermore, when} considering electrons just within the estimated scattering width, the intensity shows not only new beam electrons but stacked beam intensities including components from previous beams. However, under the assumption of a continuous scattering process and the subtraction of isotropic background intensities, the narrow width around the magnetic field direction should be primarily represented by the newest beam.
For these reasons, even with the true scattering width, the calculated energy flux is expected to be larger than the polar auroral mean.
In this study, we cannot say how strong these effects are and how much larger the energy fluxes should be. 
Nevertheless, we assume that the energy fluxes within these scattering widths, which yield values closest to the auroral mean, serve as our most reliable estimations to support the hypothesis that middle magnetosphere beams originate from upward accelerated auroral electrons.

\subsection{Relation between Intensity and Beamness}
\begin{figure}[h!]
\noindent\includegraphics[width=1\textwidth]{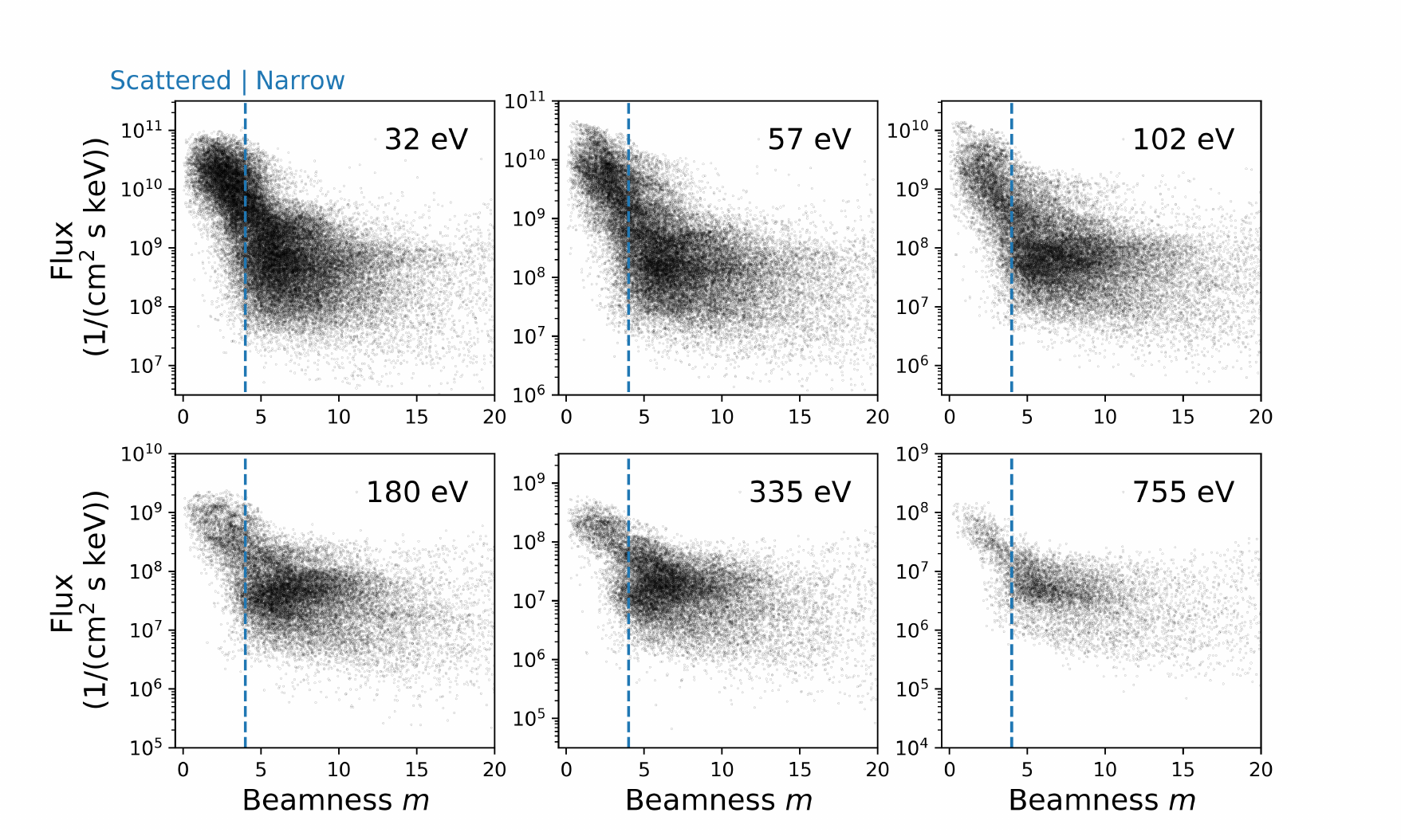}
\caption{Beam intensity integrated to $\alpha_{max}=90\degree$ and projected onto Jupiter's upper atmosphere plotted against the beamness parameter $m$ for each energy channel.}
\label{fig:intensity_beamness}
\end{figure}

A property of the beams in the middle magnetosphere, which we analyze is their intensity in relation to their beamness parameter $m$.
Figure \ref{fig:intensity_beamness} shows for each energy channel the integrated intensity or flux of the beam as a function of the beamness parameter $m$. The integrated intensity is calculated as the total integrated beam intensity with $\alpha_{max}=90\degree$ similar to Equation \ref{eq:energyflux} and is projected onto Jupiter's upper surface according to Equation \ref{eq:projection}. This allows for a more meaningful comparison between the intensities. For the purpose of this plot, we excluded the fitting results with a beamness value equal to the boundary of the fitting parameter, $m\approx0.01$.
For all energy channels, it can be clearly observed that with increasing intensity the beams show, in general, a smaller beamness $m$, indicating a larger beam width. The portion with the highest intensities is only represented by scattered beams with $m<4$, while for the lowest intensities only narrow beams with $m\geq 4$ are present. The plots demonstrate the absence of beams with relatively high intensity and large beamness $m$, as well as beams with low intensity and small beamness.
This suggests that beams with higher intensities are more scattered.

The absence of low-intensity scattered beams can be attributed to some extent to the lower detectability of stronger scattered beams and lower-intensity beams previously discussed. The latter can be seen in the figure as the intensity and therefore the total number of beams decreases with increasing energy channel.
However, it is possible that there is a physically explicable absence of strongly scattered beam measurements with low intensities. Once the beam is trapped for a sufficient amount of time to be scattered to this extent, new beams are probably already generated on the same $M$-shell, which results in an increased total intensity when these multiple beams are measured in the same time interval.

There seems to be no technical reason why narrow beams with high intensity should not be measurable.
The possible reason for the low beamness of high intensity beams involves the relationship between the degree of scattering and the lifetime of the beam.
Broad scattered beams as defined with a beamness $m<4$ may contain multiple older scattered and also new narrow beams, while in the case of narrow beams, the width limitation precludes the presence of old beams. Consequently, there may be only one or a few new ones since the time in which more new ones are generated and can increase the measured intensity is limited until the beams are scattered to strong. This allows for a greater amount of beams for the scattered beam distribution resulting in larger measured intensities.

In addition, it might be possible that there is a physical mechanism that depends on the intensity of the beam, resulting in actual increased scattering for beams with higher intensity.
It has been discussed that energetic electron beams can be the source of whistler mode waves in the Earth's magnetosphere \cite{ergun2003} and Jupiter's polar cap \cite{tetrick2017,elliott2020}. These waves may subsequently scatter the beam itself, as has been shown for the waves in the polar cap, which contribute to the pitch angle scattering of electron beams \cite{elliott2018a}. If a higher intensity beam generates stronger waves which are capable of stronger pitch angle scattering of the beam, then this may be a mechanism that could contribute to the observed trend of lower beamness for high-intensity beams. As demonstrated by \citeA{ma2024}, the growth rate of whistler waves generated by pancake electron distributions, which exhibit opposite anisotropy with respect to beams, is high for high electron intensities.

%alltenartive kann es auch ein weiterer hinweis darauf sein dass die vordereb gescatterten da sie auch andere intensitäten ausweifen von wo anders her kommen
\clearpage
\section{Conclusion}
\label{conclusion}

Our study presents an investigation of energetic electrons measured by JEDI on orbits 8 to 30 within radial distances between 13 and 50.5 $\mathrm{R_J}$. The objective is to qualitatively detect electron beams and to compare their characteristics with auroral electrons, to determine if middle magnetosphere beams are accelerated in the low-altitude auroral region, as previously anticipated.

The occurrence of bidirectional electron beams in the Jovian middle magnetosphere has previously been documented in investigations of Galileo \cite{frank2002,tomas2004a,tomas2004b,mauk2007} and Juno \cite{ma2021} electron data.
The presence of these beams on field lines that map to regions of main auroral emissions, along with their narrow width, suggests their acceleration in the low-altitude auroral region. 
The upward acceleration of electrons in the acceleration regions above the main aurora requires a mechanism that contradicts the previous expectations of only electrostatic planetward acceleration on these magnetic field lines.
This was already suggested on the basis of Galileo observations \cite{mauk2007} and further supported by Juno observations of bidirectional and broadband auroral electrons, indicative of stochastic acceleration \cite{mauk2017a,mauk2020,clark2017,saur2006,saur2018,salveter2022}.
These observations are consistent with the hypothesis that the electron beams observed in the middle magnetosphere are indeed upward-accelerated auroral electrons.
We perform a comparison between the electron beams of our study and the upward and bidirectional electrons analyzed above the main aurora in the statistical study of \citeA{salveter2022}.
The detection method shows that true narrow beams, as defined by \citeA{mauk2007}, are present throughout the observation range in the middle magnetosphere. 
The electron measurement positions in our study are estimated to magnetically map to $L$-shells of $11-15$ of the polar observation range of \citeA{salveter2022}.
For this range the polar study shows a monotonic increase of upward and bidirectional distributions relative to an increasing amount of field-aligned electron distribution.
The occurrence of narrow beams in our study shows a similar increase with increasing distance, which is consistent with the investigated hypothesis. % that the beams represent upward accelerated auroral electrons.
Based on the statistical study of auroral electrons and the estimated mapping, we hypothesize that the number of electron beams will continue to increase with greater distances. Future investigations of electrons at greater distances may reveal whether the auroral electron distribution is further observed in measurements in the middle magnetosphere and contribute to the mapping of main auroral emission to the equatorial magnetosphere.

The energy fluxes of the narrow beams are calculated for different approaches and compared with the energy fluxes of the auroral field-aligned electrons in \citeA{salveter2022}.
%which show mean values of $1-3\times10^2~\mathrm{mW~m^{-2}}$for the covered region.
Considering the total beam intensity, the energy fluxes are two to three orders of magnitude larger than the mean auroral energy fluxes. This indicates that even if only narrow beams are considered, not only newly generated beams are measured but a superposition of new beams and older trapped beams.
The energy flux of the beams just within the loss cone is more than one magnitude smaller than the auroral mean, which suggests that the majority of the beam electrons are scattered out of the loss cone and will stay trapped in Jupiter's magnetosphere.
This finding provides further support for the hypothesis that auroral electron acceleration may be a significant source of energetic electrons into Jupiter's middle magnetosphere, as discussed by \citeA<e.g.>{saur2018} and recently by \citeA{mauk2024} considering observations of upward auroral electrons with energies exceeding $10$ s of MeV.
The energy fluxes within the estimated scattering widths, as calculated by solving the pitch angle diffusion equation, exhibit values that lie between the total beam and loss cone energy fluxes.
The auroral mean energy flux lies between the energy fluxes within the scattering widths calculated with the two sets of diffusion coefficients $D_{\mathrm{Li}}$ and $D_{\mathrm{WM}}$.
It is our interpretation that the similar energy fluxes further support the hypothesis that the middle magnetosphere beams are generated in the high-latitude auroral acceleration regions and are scattered during their propagation to the middle magnetosphere.

The acceleration mechanism of the auroral electrons remains a subject of ongoing discussion. While auroral electrons with mono-energetic signatures appear to be accelerated by electrostatic potentials, wave-particle interaction is believed to be the mechanism that energizes the predominant electrons with stochastic acceleration signatures. Alfvén waves, which are observed as magnetic field fluctuations in the middle magnetosphere and sometimes in the polar magnetosphere, are being discussed to play a critical role in the acceleration process \cite{saur2003,saur2018,sulaiman2022,salveter2025}.

We performed a study on the bidirectionality of narrow electron beams to determine whether they are consistently detected simultaneously in the parallel and anti-parallel directions of the magnetic field, as described in previous observations \cite{frank2002,tomas2004a,mauk2007}.
Our results demonstrate a higher detection rate of bidirectional electron beams compared to monodirectional ones. The study showed that the detection of monodirectional beams could often be attributed to asymmetrical pitch angle coverage parallel and anti-parallel to the magnetic field direction. The results suggest that the generation of auroral electron beams may frequently occur simultaneously in both hemispheres.

The broader, scattered beams may be auroral beams, but could partially originate from adiabatic transport mechanisms.
The number of scattered beams decreases with increasing distance, starting from a maximum around $M = 23$. This inverse trend, compared to narrow beams, may be caused by inward adiabatic transport of auroral beams, which has the capacity to broaden a bidirectional beam \cite{clark2014}.
It could also be attributed to the increase in plasma transport velocity with increasing distance, as demonstrated by \citeA{bagenaldelamere2011}.
Beams generated at smaller $M$-shells could remain at these $M$-shells longer and undergo a greater degree of pitch angle scattering before being transported out of the middle magnetosphere than beams generated at larger $M$-shells. 
It is unclear whether it is possible to distinguish an adiabatically energized, broad, field-aligned distribution from a scattered beam distribution showing pitch angle scattered auroral beams based on pitch angle distributions alone.
Further research on the characteristics of scattered beams and the evolution of auroral beams is necessary to better understand their generation process. 

\noindent We summarize our results in the following list: 
\begin{enumerate}\itemsep1em
\item Narrow electron beams are present throughout the equatorial middle magnetosphere within 14 and 50 $\mathrm{R_J}$ radial distances and show an increasing occurrence with increasing distance.

 \item Based on occurrence and energy fluxes, these beams are consistent with originating from auroral upward accelerated electrons.

 \item The energy flux within the loss cone indicates that most of the beam electrons are scattered out of the loss cone and serve as an energetic electron source in the middle magnetosphere.

 \item Bidirectional narrow beams are detected more frequently than monodirectional beams, suggesting that they are often generated simultaneously in both hemispheres.

 \item Scattered electron beams show a decreasing occurrence with increasing distance. These broader beams may be pitch angle scattered auroral beams, but could partially also originate from adiabatic transport mechanisms. 

 \end{enumerate}

\appendix
% \section{Calculation of the Energy Flux}
\remove{Appendix A Calculation of the Energy Flux}

\remove{The energy flux equation as a function of the measured intensity $I$ can be derived from the particle flux definition with the phase space distribution function $f(\mathbf{v})$.
The particle flux $F_N$ through a surface is defined as the number density $n$ times the velocity component perpendicular to the surface $v_N$ and can be expressed by the first-order velocity moment} 

% \cite[p. 121, 125]{baumjohann}.
% \begin{linenomath*}
%     \begin{equation}
%         F_N =\int d^3v~f( \mathbf{v})~ \mathbf{v_N}
%     \end{equation}
% \end{linenomath*}
\remove{For a particle flux through a surface perpendicular to the magnetic field direction, $v_N$ is equal to the parallel velocity component $\mathbf{v_\parallel}=v\cos{\alpha} $ of the particles. }

% \begin{linenomath*}
% \begin{eqnarray}
% F_N  &&= \int d^3v~f( v_\parallel,v_\perp,\alpha)~v~\cos\alpha\\
%     &&=\int_{\theta_{min}}^{\theta_{max}} d\theta\int_{\alpha_{min}}^{\alpha_{max}}d\alpha\int_{v_{min}}^{v_{max}}dv~f(v_\parallel,v_\perp,\alpha)~v^3~    \cos\alpha\sin\alpha
% \end{eqnarray}
% \end{linenomath*}
\remove{where $v_\perp=v\sin\alpha$ is the perpendicular velocity component of the particles.
The measured intensity or differential particle flux $I(E,\alpha)$ as a function of energy can be related to the distribution function by $f(\mathbf{v},\alpha )~v^3~dv~d\Omega = I(E,\alpha)~dE~d\Omega$ as described in }
% \citeA[p. 121]{baumjohann}, 
\remove{leading to}

% \begin{linenomath*}
% \begin{equation}
% F_N=\int_{\theta_{min}}^{\theta_{max}} d\theta\int_{\alpha_{min}}^{\alpha_{max}} d\alpha\int_{E_{min}}^{E_{max}} dE~I(E,\alpha ) \cos\alpha~\sin\alpha
% \end{equation}
% \end{linenomath*}
\remove{The differential energy flux can be obtained by multiplying the differential particle flux by the particle energy. 
Since the intensity is measured and fitted within specific energy channels $E_i$ within the limits $E_{i,min}$ and $E_{i,max}$, these particles are assigned to the mean energy of that channel $\bar{E_i}$. 
The calculation of the energy flux of each energy channel is simplified to}

% \begin{linenomath*}
% \begin{equation}
%     F_{E_i} = \bar{E_i}\cdot \Delta E \int_{\theta_{min}}^{\theta_{max}} d\theta\int_{\alpha_{min}}^{\alpha_{max}} d\alpha~I_i( \alpha )  ~\cos\alpha~\sin\alpha
%     \label{eq:singlefe}
% \end{equation}
% \end{linenomath*}
 \remove{where $I_i( \alpha )$ is the beamness function fit of the energy channel $E_i$. The central energy $\bar{E_i}$ and the width of the energy channels $\Delta E_i$ of the energy channel are calculated by}
 
%  \begin{linenomath*}
% \begin{eqnarray}
%      &&\bar{E_i}=\sqrt{E_{i,min}\cdot E_{i,max}}\\
%      &&\Delta E_i = E_{i,max}- E_{i,min}
%  \end{eqnarray}
%  \end{linenomath*}
\remove{The total energy flux $F_E$ can be calculated as the sum of the energy fluxes $F_{E_i}$ of each channel $F_E = \sum  F_{E_i}$. 
Since we are interested in the energy flux of the beam, only the intensity above the isotropic background intensity is used. Therefore, the isotropic background intensity, defined by the parameter $B_i$, is subtracted from the intensity fit $I_i$. The intensity is integrated over the entire azimuthal angle assuming gyrotropicity. The energy flux equation for each energy channel $i$ is then given by}

\section{Cost Function Space Analysis}

\begin{figure}[h!]
\noindent\includegraphics[width=1\textwidth]{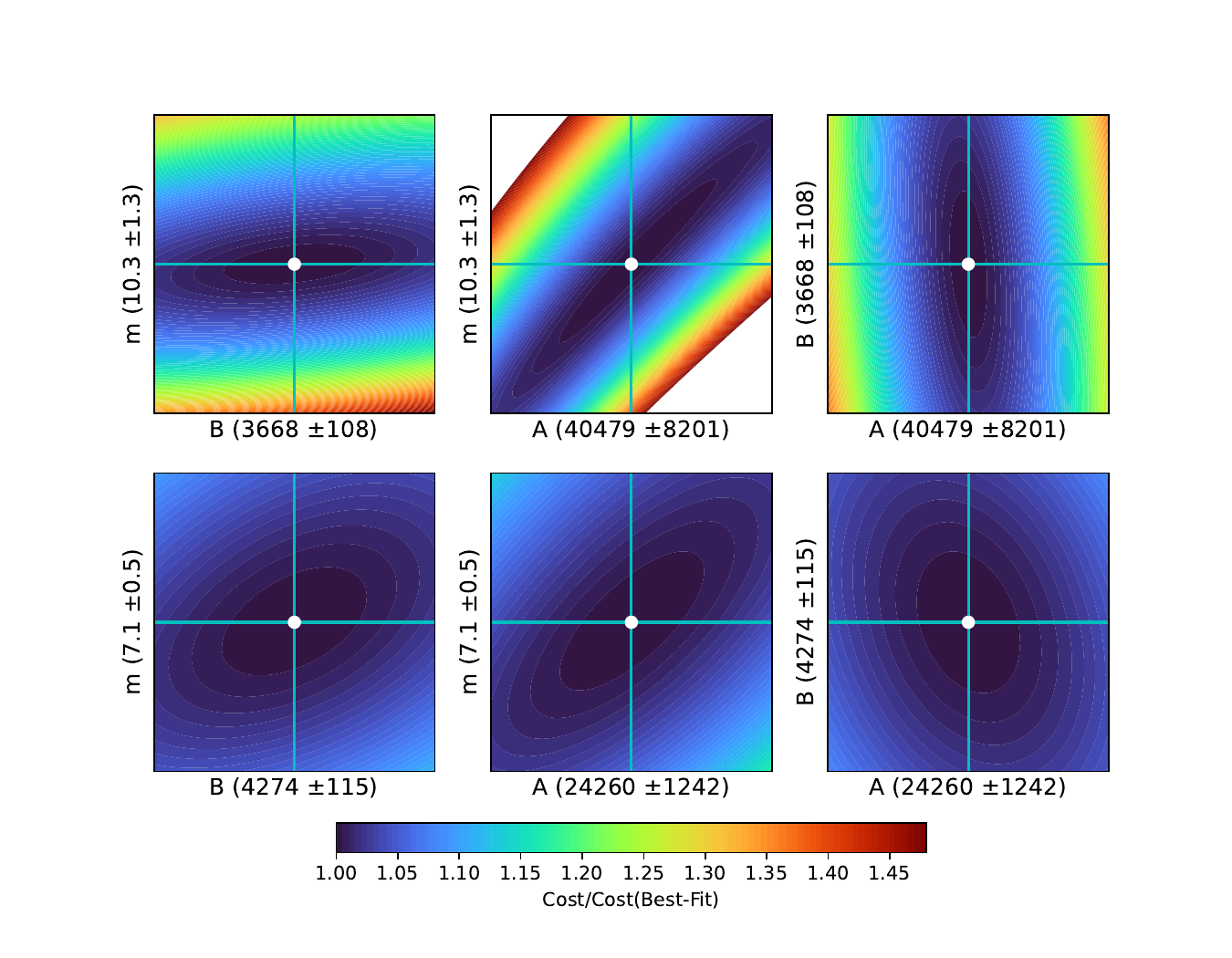}
\caption{Cost function variations as a function of the fitting parameters $A$, $m$, and $B$. The displayed values show the cost function relative to the cost function value of the best fit parameters which are marked as \change{black}{white} dots, within $\pm \sigma$ of the best fit parameters. The upper plots show the results for 01:38 UTC and the lower plots show the results for 01:46 UTC at day 88 of the year 2018.}
\label{fig:cost}
\end{figure}

An optimized fit can be validated by analyzing the cost function for varying values of the fitting parameters.
The cost function $\mathrm{Cost} = 0.5\cdot \sum_n^N(I_{data}(\alpha_n)-I_{fit}(\alpha_n))^2$ describes the difference between the data $I_{data} $ and the fitted beam function $I_{fit}$ for $N$ data points and is minimized in the fitting algorithm.
A successful fit yields optimized parameters that correspond to the global minimum of the cost function within the parameter space, as described in \citeA{kim2020}.
Figure \ref{fig:cost} shows the variations of the cost function as a function of the fitting parameters $A$, $B$, and $m$ of the example measurement intervals. The displayed values show the cost function relative to the cost function value of the best fit parameters, within $\pm \sigma$ of the best fit parameters. The upper plots show the results for 01:38 UTC and the lower plots show the results for 01:46 UTC at day 88 of the year 2018.
These plots show that the best-fit parameters result in a minimum cost function value, demonstrating that the fitting minimization was successful.
The results reveal a positive correlation between the fitting parameters $A$ and $m$, with the degree of correlation being influenced by the coverage of the pitch angle. 

%%%%%%%%%%%%%%%%%%%%%%%%%%%%%%%%%%%%%%%%%%%%%%%
% Optional Glossary, Notation or Acronym section goes here:
%
% Glossary is only allowed in Reviews of Geophysics
%  \begin{glossary}
%  \term{Term}
%   Term Definition here
%  \term{Term}
%   Term Definition here
%  \term{Term}
%   Term Definition here
%  \end{glossary}

%%%%%%%%%%%%%%%%%%%%%%%%%%%%%%%%%%%%%%%%%%%%%%%
% Acronyms
%% NOTE that acronyms in the final published version will be spelled out when used in figure captions.
%   \begin{acronyms}
%   \acro{Acronym}
%   Definition here
%   \acro{EMOS}
%   Ensemble model output statistics
%   \acro{ECMWF}
%   Centre for Medium-Range Weather Forecasts
%   \end{acronyms}

%%%%%%%%%%%%%%%%%%%%%%%%%%%%%%%%%%%%%%%%%%%%%%%
% Notation
%   \begin{notation}
%   \notation{$a+b$} Notation Definition here
%   \notation{$e=mc^2$}
%   Equation in German-born physicist Albert Einstein's theory of special
%  relativity that showed that the increased relativistic mass ($m$) of a
%  body comes from the energy of motion of the body—that is, its kinetic
%  energy ($E$)—divided by the speed of light squared ($c^2$).
%   \end{notation}

%%%%%%%%%%%%%%%%%%%%%%%%%%%%%%%%%%%%%%%%%%%%%%%
%
% DATA SECTION and ACKNOWLEDGMENTS
%
%%%%%%%%%%%%%%%%%%%%%%%%%%%%%%%%%%%%%%%%%%%%%%%
%TC:ignore

\section*{Conflict of Interest Statement}
The authors have no conflicts of interest to disclose.

\section*{Open Research Section}
The calibrated JEDI data used are available from the website of the NASA Planetary Data System: Planetary Plasma Interactions (https://pds-ppi.igpp.ucla.edu/collection/JNO-J-JED-3-CDR-V1.0) \cite{jedidata}. The Python wrapper of C++ of the community coding project \cite{wilson2023, James_JupiterMag} is utilized to calculate the $M$-shell parameter based on the Jovian internal magnetic field model JRM33 \cite{connerney2022} and the magnetodisc current sheet model CON2020 \shortcite{connerney2020}.

\acknowledgments
This research was supported by the Graduate School of Geosciences (GSGS) of the University of Cologne, NASA grant 80NSSC20K0560 and NASA Juno contract NNM06AA75C.

%TC:endignore
%%%%%%%%%%%%%%%%%%%%%%%%%%%%%%%%%%%%%%%%%%%%%%%
% REFERENCES and BIBLIOGRAPHY
%
% \bibliography{<name of your .bib file>} don't specify the file extension
% don't specify bibliographystyle
%
%%%%%%%%%%%%%%%%%%%%%%%%%%%%%%%%%%%%%%%%%%%%%%%
\clearpage
\bibliography{piaseckibib}
%
%Reference citation instructions and examples:
%
% Please use ONLY \cite and \citeA for reference citations.
% \cite for parenthetical references
% ...as shown in recent studies (Simpson et al., 2019)
% \citeA for in-text citations
% ...Simpson et al. (2019) have shown...
%
%
%...as shown by \citeA{jskilby}.
%...as shown by \citeA{lewin76}, \citeA{carson86}, \citeA{bartoldy02}, and \citeA{rinaldi03}.
%...has been shown \cite{jskilbye}.
%...has been shown \cite{lewin76,carson86,bartoldy02,rinaldi03}.
%... \cite <i.e.>[]{lewin76,carson86,bartoldy02,rinaldi03}.
%...has been shown by \cite <e.g.,>[and others]{lewin76}.
%
% apacite uses < > for prenotes and [ ] for postnotes
% DO NOT use other cite commands (e.g., \citet, \citep, \citeyear, \nocite, \citealp, etc.).
\end{document}